\documentclass[runningheads]{llncs}
\usepackage{thm-restate}
\usepackage{hyperref}

\usepackage{amsmath,amssymb}
\usepackage{stmaryrd}
\usepackage{todonotes}
\usepackage{cite}
\usepackage{tcolorbox}
\usepackage{xspace}
\usepackage{pifont}
\usepackage{mathpartir}
\usepackage[capitalise]{cleveref}
\usepackage{thm-restate}
\usepackage{url}
\usepackage{xspace}
\usepackage{subcaption}
\usepackage{listings}
\usepackage{makecell}
\usepackage{subcaption}
\usepackage{tablefootnote}
\usepackage{tikz}
\usetikzlibrary{positioning, shapes, arrows.meta, fit, calc, shapes.symbols, shapes.multipart, bbox}
\usepackage[inline]{enumitem}
\usepackage{booktabs}
\usepackage[misc]{ifsym}

\usepackage{ifthen}
\newboolean{hideAppendix}
\setboolean{hideAppendix}{false}

\expandafter\def\expandafter\normalsize\expandafter{%
    \normalsize%
    \setlength\abovedisplayskip{5pt}%
    \setlength\belowdisplayskip{5pt}%
}

\def\orcidID#1{\smash{\href{http://orcid.org/#1}{\protect\raisebox{-1.25pt}{\protect\includegraphics{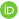}}}}}

\newcommand*\circled[1]{\tikz[baseline=(char.base)]{
            \node[shape=rectangle,fill,inner sep=2pt] (char) {\textcolor{white}{#1}};}}

\newcommand{\Sometimes}{\cmark\xspace}
\newcommand{\Never}{\xmark\xspace}
\newcommand{\Crash}{Unsup.\xspace}

\newcommand{\numOtherTests}{69\xspace}
\newcommand{\numTestsTotal}{169\xspace}
\newcommand{\ToolName}{\texttt{jMT}\xspace}
\newcommand{\jcstress}{\textsc{jcstress}\xspace}
\newcommand{\JavaSim}{\textsc{JavaSim}\xspace}
\newcommand{\herd}{\textsc{herd}\xspace}
\newcommand{\cat}{\texttt{.cat}\xspace}
\newcommand{\jamtwentyone}{\ensuremath{\text{\texttt{JAM}}_{21}}\xspace}
\newcommand{\jls}{\ensuremath{\text{\texttt{JLS}}_{04}}\xspace}
\newcommand{\VarHandle}{\texttt{VarHandle}\xspace}
\newcommand{\LBx}{\texttt{LbOdd}\xspace}

\newcommand{\ALT}{\;\;|\;\;}
\newcommand{\sep}{\;\kw{;}\;}

\newcommand{\ctid}[1]{\mathtt{T}_#1}
\newcommand{\cloc}[1]{\mathtt{#1}}
\newcommand{\creg}[1]{\mathtt{#1}}
\newcommand{\sexpr}{\ensuremath{sxp}}
\newcommand{\sexprr}{\ensuremath{sxp_\lR}}
\newcommand{\sexprw}{\ensuremath{sxp_\lW}}
\newcommand{\expr}{\ensuremath{exp}}

\newcommand{\eqdef}{\triangleq}
\renewcommand{\iff}{\Leftrightarrow}

\newcommand{\set}[1]{\ensuremath{\{#1\}}}

\newcommand{\textcode}[1]{{\texttt{#1}}}
\newcommand{\kw}[1]{\textbf{\textcode{#1}}}
\newcommand{\inlineheadingbf}[1]{\medskip\noindent{\bfseries #1.}}

\newcommand{\makeset}[1]{\ensuremath{\mathit{#1}}}

\newcommand{\tid}{\ensuremath{\tau}}
\newcommand{\TIDs}{\ensuremath{\mathsf{TID}}}
\newcommand{\val}{\ensuremath{v}}

\newcommand{\Val}{\ensuremath{\mathsf{Val}}}
\newcommand{\reg}{{a}}
\newcommand{\Reg}{\ensuremath{\mathsf{Reg}}}
\newcommand{\loc}{{x}}
\newcommand{\loca}{{y}}

\newcommand{\Loc}{\ensuremath{\mathsf{Loc}}}

\newcommand{\Lab}{\ensuremath{\mathsf{Lab}}}
\newcommand{\evt}{e}
\newcommand{\Events}{\ensuremath{\mathsf{Events}}}
\newcommand{\ini}{\ensuremath{\mathsf{ini}}}

\newcommand{\cmark}{\ding{51}}
\newcommand{\xmark}{\ding{55}}

\colorlet{colorDEPS}{violet}
\colorlet{colorPO}{darkgray!80!black}
\colorlet{colorRF}{green!60!black}
\colorlet{colorMO}{orange}
\colorlet{colorFR}{purple}
\colorlet{colorECO}{red!80!black}
\colorlet{colorSYN}{violet} 
\colorlet{colorHB}{blue}
\colorlet{colorSO}{magenta}
\colorlet{colorPPO}{magenta}
\colorlet{colorPB}{olive}
\colorlet{colorRMW}{pink!70!black}
\colorlet{colorRSEQ}{blue}
\colorlet{colorSC}{violet}
\colorlet{colorPSC}{violet}
\colorlet{colorSCB}{violet}
\colorlet{colorREL}{blue!65!green}
\colorlet{colorACQ}{red!80!green}
\colorlet{colorCONFLICT}{olive}
\colorlet{colorRACE}{olive}

\newcommand{\rlab}[3]{{\lR}^{#1}({#2},{#3})}
\newcommand{\wlab}[3]{{\lW}^{#1}({#2},{#3})}
\newcommand{\rmwlab}[4]{{\lRMW}^{#1}({#2},{#3},{#4})}
\newcommand{\flab}[1]{{\lF^{#1}}}

\newcommand{\lR}{{\sR}}
\newcommand{\lW}{{\sW}}

\newcommand{\lRMW}{{\sRMW}}
\newcommand{\lF}{{\sF}}

\newcommand{\lTYP}{{\mathtt{typ}}}
\newcommand{\lLOC}{{\mathtt{loc}}}

\newcommand{\lTID}{{\mathtt{tid}}}

\newcommand{\lSXPR}{{\mathtt{sxp}_\lR}}
\newcommand{\lSXPW}{{\mathtt{sxp}_\lW}}

\newcommand{\sR}{\mathsf{R}}
\newcommand{\sW}{\mathsf{W}}
\newcommand{\sRMW}{\mathsf{RMW}}
\newcommand{\sF}{\mathsf{F}}

\newcommand{\pcmd}{c}

\newcommand{\cmd}{C}

\newcommand{\skipc}{\kw{SKIP}}
\newcommand{\readInst}[3]{#1 \;{:=}\;\readInstn^{#3}({#2})}
\newcommand{\writeInst}[3]{\writeInstn^{#3}(#1,#2)}
\newcommand{\assignInst}[2]{#1\;{:=}\;#2}

\newcommand{\casInst}[5]{#1 \;{:=}\;\casInstn^{#5}({#2},{#3},{#4})}
\newcommand{\fenceInst}[1]{\kw{fence}^{#1}}
\newcommand{\writeInstn}{\kw{STORE}}

\newcommand{\readInstn}{\kw{LOAD}}
\newcommand{\casInstn}{\kw{CAX}}

\newcommand{\ite}[3]{\kw{if}\;#1\:\kw{then}\;#2\; \kw{else}\;#3}

\newcommand{\rmd}{\ensuremath{\mathsf{rm}}}

\newcommand{\wmd}{\ensuremath{\mathsf{wm}}}

\newcommand{\fmd}{\ensuremath{\mathsf{fm}}}

\newcommand{\mRexpr}{\ensuremath{\mathit{rexpr}}}
\newcommand{\mWexpr}{\ensuremath{\mathit{wexpr}}}

\newcommand{\Executions}{\ensuremath{\mathcal{G}}}
\newcommand{\Committed}{\makeset{C}}
\newcommand{\Constraints}{\ensuremath{\Gamma}}

\newcommand{\Reads}{\makeset{R}}
\newcommand{\Writes}{\makeset{W}}

\newcommand{\makerel}[2][.]{\ensuremath{{\color{#1}\mathsf{#2}}}}

\newcommand{\seqR}[2]{\ensuremath{#1;#2}}

\newcommand{\transC}[1]{\ensuremath{#1}^{+}}
\newcommand{\reftransC}[1]{\ensuremath{#1}^{*}}
\newcommand{\transred}[1]{\ensuremath{#1}^{-}}

\newcommand{\inv}[1]{\ensuremath{#1^{-1}}}

\newcommand{\tup}[1]{{\langle{#1}\rangle}}

\newcommand{\rst}[1]{|_{#1}}

\renewcommand{\int}{\makerel{int}}

\newcommand{\withthread}[2]{\ensuremath{#1 \rst{#2}}}

\newcommand{\po}{\makerel[colorPO]{po}}

\newcommand{\rf}{\makerel[colorRF]{rf}}

\newcommand{\invrf}{\inv{\rf}}

\newcommand{\coe}{\coe}

\newcommand{\sw}{\makerel[colorSYN]{sw}}

\newcommand{\hb}{\makerel[colorHB]{hb}}

\newcommand{\so}{\makerel[colorSO]{so}}

\tikzset{
   every path/.style={>=stealth},
   po/.style={->,color=colorPO,thin,shorten >=-0.5mm,shorten <=-0.5mm},
   sw/.style={->,color=colorSYN,shorten >=-0.5mm,shorten <=-0.5mm},
   rf/.style={->,color=colorRF,thin,shorten >=-0.5mm,shorten <=-0.5mm},
   hb/.style={->,color=colorHB,thick,shorten >=-0.5mm,shorten <=-0.5mm},
   mo/.style={->,color=colorMO,dotted,very thick,shorten >=-0.5mm,shorten <=-0.5mm},
   no/.style={->,dotted,thick,shorten >=-0.5mm,shorten <=-0.5mm},
   fr/.style={->,color=colorFR,dotted,thick,shorten >=-0.5mm,shorten <=-0.5mm},
   deps/.style={->,color=colorDEPS,dotted,thick,shorten >=-0.5mm,shorten <=-0.5mm},
   vis/.style={->,color=colorDEPS,dotted,thick,shorten >=-0.5mm,shorten <=-0.5mm},
   rmw/.style={->,color=colorRMW,thick,shorten >=-0.5mm,shorten <=-0.5mm},
   ppo/.style={->,color=colorPPO,thick,shorten >=-0.5mm,shorten <=-0.5mm},
}

\newcommand{\readvalue}[1]{\ensuremath{{\color{colorRF}#1}}}
\newcommand{\itreads}[2][/\!\!/]{\ensuremath{{\color{colorRF}\;#1 \readvalue{#2}}}}

\newcommand{\allowed}{{\color{green}\cmark}}
\newcommand{\forbidden}{{\color{red}\xmark}}
\newcommand{\lPln}{\ensuremath{\mathbf{pln}}}
\newcommand{\lOpq}{\ensuremath{\mathbf{opq}}}
\newcommand{\lRel}{\ensuremath{{\color{colorREL}\mathbf{rel}}}}
\newcommand{\lAcq}{\ensuremath{{\color{colorACQ}\mathbf{acq}}}}
\newcommand{\lVol}{\ensuremath{{\color{colorSO}\mathbf{vol}}}}

\newcommand{\lStoreStore}{\ensuremath{\lW\lW}}
\newcommand{\lLoadLoad}{\ensuremath{\lR\lR}}
\newcommand{\lFull}{\ensuremath{{\color{colorSO}\mathbf{full}}}}

\newcommand{\rdx}[1]{\lR #1}
\newcommand{\wrx}[1]{\lW #1}

\definecolor{codegreen}{rgb}{0.02 0.5 0.09}
\definecolor{codegray}{rgb}{0.46,0.48,0.54}
\definecolor{codeblue}{rgb}{0,0.2,0.7}
\definecolor{codebg}{rgb}{0.96,0.965,0.97}
\definecolor{codefg}{rgb}{0.03,0.03,0.03}
\lstdefinestyle{mystyle}{
	backgroundcolor=\color{codebg},
	commentstyle=\color{codegreen},
	keywordstyle=\color{codeblue},
	numberstyle=\tiny\color{codegray},
	stringstyle=\color{codegreen},
	basicstyle=\ttfamily\footnotesize\color{codefg},
	breakatwhitespace=false,
	showstringspaces=false,
	numbers=left,
}
\lstdefinelanguage{cat}
{morekeywords={let,with,from,empty,irreflexive,acyclic,show},
sensitive=false,
morecomment=[s]{(*}{*)},
morestring=[b]",
}

\title{\ToolName: Testing Correctness of Java Memory Models}
\author{
	Lukas Panneke\textsuperscript{(\Letter)\,\orcidID{0009-0008-9241-5583}}
	\and Heike Wehrheim\textsuperscript{\,\orcidID{0000-0002-2385-7512}}}
\institute{
	Carl von Ossietzky Universität Oldenburg, Oldenburg, Germany\\
	\email{\{lukas.panneke,heike.wehrheim\}@uol.de}
}
\authorrunning{L.\ Panneke \ and H.\ Wehrheim}

\begin{document}
\maketitle

\begin{abstract}
According to common belief,  the Java memory model is broken. In the past, several approaches have proposed repairs,
 often only to find new programs exhibiting unexpected, unintuitive behavior or the model forbidding standard compiler optimizations.
 The complexity of defining a memory model for concurrent Java lies in the fact that it requires a {\em multi-execution model}.
 Multi-execution models need to inspect many potential executions of a program in order to find the valid ones.
 Tools automatically validating novel proposals of Java memory models are, however, largely lacking.

To alleviate this problem, we introduce \ToolName, a novel tool for constructing multi-execution semantics for
concurrent Java programs. \ToolName relies on single-execution models defining {\em well-formed} execution graphs,
based on which it builds valid multi-execution semantics via causality checking.
Thereby, \ToolName supports evaluating new proposals of Java memory models (JMMs) on a per-program basis.
\ToolName can furthermore be employed for testing the conformance of JMMs to existing compilation schemes and compilers.
Our evaluation of \ToolName on \numTestsTotal litmus tests reveals a number of interesting insights into existing JMMs.
\end{abstract}


\section{Introduction} \label{sec:intro}
In the past,
there have been numerous efforts for correctly defining the Java Memory Model (JMM)~\cite{
	DBLP:conf/popl/MansonPA05,
	DBLP:journals/pacmpl/BenderP19,
	DBLP:conf/ecoop/LiuBP22
},
i.e., defining the semantics of concurrent Java programs on multiprocessor machines.
The official 1996 specification~\cite{Javaspec}
has been shown to be flawed by Pugh~\cite{DBLP:conf/java/Pugh99}.
The next 2004 proposal of a JMM (Manson et al.~\cite{DBLP:conf/popl/MansonPA05},
in the sequel called \jls)
has also been criticized
for being incompatible with some desired compiler optimizations~\cite{
	DBLP:conf/esop/CenciarelliKS07,
	ecoop/SA08
}.
Moreover, it is outdated as far as modern synchronization primitives are concerned,
in particular the features of the \VarHandle API.
The \VarHandle API adds
new access modes for variables (opaque and release/acquire),
read-modify-write operations,
memory fences
and the mixing of volatile and non-volatile accesses to the same location.
The intended semantics of this API has been extensively described
by its designer Doug Lea~\cite{j9mm}.
The most recent JMM proposal
of Bender, Liu and Palsberg~\cite{
	DBLP:journals/pacmpl/BenderP19,
	DBLP:conf/ecoop/LiuBP22
}
(in the following \jamtwentyone)
incorporates most of these new features,
the proposed model -- however --
also forbids certain expected, observable behavior,
and allows undesirable behavior,
as we show below.

The core reason for this difficulty in defining a correct JMM
lies in the fact that Java
requires a {\em multi-execution model}:
the validity of executions
(i.e., whether some particular execution and its outcome should be allowed or not)
cannot be determined by looking at one execution in isolation.
Instead, given a set of candidate executions,
a subset of valid executions is to be constructed.
Java memory models therefore often consist of two parts:
a single-execution model (SEM)
defining a set of candidates
and on top of that a multi-execution model (MEM)
used for validating the executions in the candidate set.
As Batty et al.~\cite{DBLP:conf/esop/BattyMNPS15}
have discussed within the context of C++,
single-execution models alone can never suffice
without restricting compiler optimizations.
Basically, the reason for this is the so called ``thin-air-problem''
in which values (of shared locations) circularly depend on their own computation.

Given the past experiences in constructing new JMMs,
every new proposal first has to undergo several healthiness checks:
(1) evaluating the novel semantics on numerous {\em litmus tests}
(small programs with interesting behavior),
(2) checking agreement with standard compilation schemes and hardware memory models, and
(3) validating the conformance of compilers with the semantics.
Currently, in particular check (2) is ultimately carried out via formal proving~\cite{
	ecoop/PetriVJ15,
	jam21,pldi/LVKHD17
}.
Proofs are, however, too labour-intensive
to be performed on every change of the model.
A tool allowing to first of all {\em test} such properties 
would enable fast iteration of model changes
so that expensive proving can be deferred to later stages of the model development,
once testing finds no further inconsistencies.

Currently, the most prominent tool supporting the development of new memory models is \herd:
\herd allows for the definition of single-execution models,
independent of any specific programming language or hardware.
\herd does, however, not support multi-execution models.
A second potentially usable tool is \JavaSim~\cite{ftfjp/MP02}
which is however known to contain bugs. 

\smallskip
\noindent In this paper,
we therefore introduce the novel tool \ToolName for multi-execution models.
\ToolName allows for  {\em testing} the above-mentioned healthiness checks
for novel Java memory models.
To this end,
a user of the tool has to develop a new single-execution model for Java
and specify it in the \cat language of \herd.
\ToolName then automatically calculates the semantics of concurrent programs,
thereby employing \ToolName's multi-execution model which builds on \jls's MEM (as this still remains the official JMM).
We provide an operational formulation of the \jls-style causality semantics,
turning its declarative definition into an executable construction of justification sequences.
Contrary to \jls, our  multi-execution model is based on the concept of {\em symbolic execution graphs}. 
Such graphs allow us to efficiently represent infinitely many candidate executions within a single graph,
and to solve parts of the justification search using an SMT solver.
Symbolic execution graphs have also in parallel been developed in~\cite{pacmpl/RichardsWCB25}, 
however, are therein employed for different forms of justifications.  

\ToolName can be used in three different scenarios:
\circled{1} to generate the semantics of litmus tests and compare
the outcomes against expected outcomes,
\circled{2} to compile programs using standard compilations schemes to hardware
and contrast the outcomes of litmus tests on the new JMM
with outcomes of the compiled program, and
\circled{3} to compile programs using standard Java compilers and then once more compare outcomes.
To support these three application scenarios,
we provide -- along with \ToolName -- a collection of \numTestsTotal litmus tests,
assembled from \cite{
	causality-test-cases,
	manson-thesis,
	jcstress,
	ecoop/SA08,
	pldi/LVKHD17,
	jam21,
	j9mm,
	jmm-jls,
	jmaonson-blog-volatile-fields-and-synchronization,
	strong-RA,
	DBLP:journals/pacmpl/BenderP19,
	shipilev-blog-close-encounters-of-jmm-kind,
	ori-against-sdep,
	unbublished-popl-special-jmm
}.
We have furthermore implemented a tiny compiler to Intel's x86 processor
to support scenario \circled{2}.
This compiler does not perform any optimizations,
so that a new JMM can be checked with respect to a compilation scheme alone.
Finally, to support scenario \circled{3}
our tool includes a toolchain
in which programs can be compiled using standard Java compilers.
For this purpose we employ a Java concurrency stress testing tool (\jcstress~\cite{jcstress})
to enable finding mismatches in behaviors at runtime.

We have implemented and experimentally evaluated \ToolName.
Our evaluation shows that \ToolName is able to find surprising novel issues with existing JMMs
and that \ToolName is
-- according to our testing and given the right SEM --
an accurate simulator for \jls.

\paragraph{Related Work.}\label{sec:related}
Several tools have been proposed
for simulation, model checking, and compilation validation
under weak memory.
\JavaSim~\cite{ftfjp/MP02} is an early simulator for the Java memory model.
It was developed alongside \jls
and predates the \texttt{VarHandle} APIs.
Consequently, it lacks support for many current language features.
Discrepancies between its results and \jls
further limit its applicability.
Many other multi-execution models have tool support as well~\cite{
	esop/PaviottiCPWOB20, 
	pacmpl/JeffreyRBCKP22, 
	pacmpl/RichardsWCB25, 
	pacmpl/MoiseenkoKV22}. 
Those tools support our scenario~\circled{1}
for their respective memory model,
instead of the Java memory model.
Many more tools target single-execution models.

\herd~\cite{herding-cats} and \textsc{Dartagnan}~\cite{dartagnan}
are prominent tools for evaluating litmus tests for \cat-specified models,
but since all \cat models are \emph{single-execution},
they cannot directly support multi-execution semantics as in~\jls.
\textsc{jmc}~\cite{concur/JorshariKMN25} and \textsc{GenMC}~\cite{GenMC}
are memory model parametric model checkers requiring acyclic $\po \cup \rf$,
excluding \jls, which does allow such cycles in the presence of data races.
They can still analyze Java programs proven race-free by other means.

We are, to our knowledge,
the first to apply testing scenarios~\circled{2} and~\circled{3}
to \emph{multi-execution} models.
\textsc{T{\'{e}}l{\'{e}}chat}~\cite{DBLP:conf/cgo/GeesonS24}
detects weak memory bugs in optimizing production compilers
using static analysis with \herd,
whereas we use a non-optimizing compiler in scenario~\circled{2}
and dynamic analysis in scenario~\circled{3}.
\textsc{TriCheck}~\cite{asplos/TrippelMLPM17}
-- similarly to \ToolName -- validates compilation schemes,
but for the \emph{single-execution} C++ model.

\paragraph{Limitations.}
Our tool supports all features of the \VarHandle API,
but lacks support for other aspects of the Java programming language
relevant to visibility and ordering:
monitors, external events (such as I/O), final fields, and
observations of thread lifetime (first and last event, start and termination detection).
We furthermore do not support \cat models with mutually recursive relations.

\paragraph{Contributions.}
Beyond tool implementation,
this paper makes the following contributions:
(1) We introduce symbolic execution graphs.
(2)  
We are (to our knowledge) the first to formalize the cross-execution event equivalence,
a key but underspecified part of \jls's MEM.
(3) We show that \jamtwentyone is flawed via counterexamples,
including mismatches between intended and actual behavior,
and a counterexample to its claimed compilation correctness to x86. 
(4) We identify a contradiction in Lea's VarHandle documentation~\cite{j9mm},
the de facto source for volatile semantics
in presence of access mode mixing on the same variable.

\section{Foundations} \label{sec:foundations}

\begin{figure}[t]
	\centering
	\small
	\begin{tabular}{rll@{\ \ \ }rll}
\emph{values} & $\,$ & $\val \in \Val = \set{0,1,\ldots}$
& \emph{locations} & $\,$ & $\loc, \loca \in \Loc = \set{\cloc{x},\cloc{y},\ldots}$
\\ \emph{local registers} & $\,$ & $\reg \in \Reg = \set{\creg{a},\creg{b},\ldots}$
& \emph{thread identifiers} & $\,$ & $\tid \in \TIDs = \set{1,2, \ldots}$ \\
\emph{read modes} & $\,$ & $\rmd \in \set{\lPln, \lOpq, \lAcq, \lVol}$ & \emph{write modes} & $\,$ & $\wmd \in \set{\lPln, \lOpq, \lRel, \lVol}$ \\
\emph{fence modes} & $\,$ & $\fmd \in \set{\lStoreStore, \lLoadLoad, \lAcq, \lRel, \lFull}$ &&&

\end{tabular}

\[\begin{array}{@{} l l @{}}
\expr ::= & \reg \ALT \val \ALT \expr + \expr \ALT \expr = \expr \ALT \expr > \expr \ALT \neg \expr \ALT \expr \land \expr \ALT \ldots
\\
\sexpr ::= & \reg \ALT \val \\
\pcmd ::= &
\assignInst{\reg}{\expr}
\ALT \writeInst{\loc}{\sexpr}{\textsf{wm}}
\ALT \readInst{\reg}{\loc}{\textsf{rm}}
\ALT \fenceInst{\textsf{fm}}
 \ALT \casInst{\reg}{\loc}{\sexpr}{\sexpr}{\textsf{rm,wm}} \\
\cmd ::= &
\skipc
\ALT \pcmd
 \ALT \cmd \sep \cmd \ALT \ite{\expr}{\cmd}{\cmd}
\end{array}
\]
\vspace{-10pt} 
\caption{Syntax of sequential programs.}
\label{fig:syntax}
\end{figure}

We start by introducing some basic definitions and notations.

\inlineheadingbf{Syntax of programs} \Cref{fig:syntax} defines the syntax of sequential programs\footnote{Alike other tools for weak memory models, our tool does not consider loops.}. We do not use Java syntax here, just plain pseudocode syntax.
Sequential programs consist of assignments to registers, store (write) and load (read) instructions, writing to and reading from shared locations as well as a read-modify-write (RMW\footnote{As RMW, we include a compare-and-exchange instruction; others are also possible.}) instruction Compare-And-eXchange. We assume programs to be in static single assignment form and registers to be disjoint between threads.  Programs might furthermore contain fences. Stores, loads, RMWs and fences come with {\em access modes} (as of the \VarHandle API).  A concurrent program $P$ is then a parallel composition (denoted $||$) of sequential programs, one for each thread in $\TIDs$. We do not give a semantics here; the semantics is fixed by a single execution model (SEM).

Execution of programs give rise to execution graphs in which the nodes are {\em events}\footnote{In \jls, these are called actions.} and edges describe relations over events.
Events are labelled over a set of labels $\Lab$.
In our multi-execution model, we build {\em symbolic} execution graphs in which the labels contain symbolic expressions $\sexpr$ (register names or constants) instead of concrete values only, i.e.,
	\begin{eqnarray*} \Lab & = & \{\rlab{\rmd}{\loc}{\sexprr}, \wlab{\wmd}{\loc}{\sexprw},\rmwlab{\rmd,\wmd}{\loc}{\sexprr}{\sexprw}, \flab{\fmd} \\
		&& \qquad \mid \sexprr, \sexprw \in \Val \cup \Reg \}\ .
	\end{eqnarray*}
The functions $\lTYP$, $\lLOC$, $\lSXPR$ and $\lSXPW$
return (when applicable) the type ($\lR/\lW/\lRMW/\lF$), location,
read   and  written symbolic expression of a given label.
An execution of an instruction by a thread $\tid \in \TIDs$ leads to an {\em event} labelled via $\Lab$ (plus with a unique identifier).
Initializing writes are specific events: They have no access modes and are performed by a special thread $\ini$ (initializing events of the form $\lW_\ini(\loc,0)$)\footnote{For simplicity, we assume initial values to be 0.}.
 Together, the set of events is $\Events = (\Lab \times \TIDs \times \mathbb{N}) \cup \set{\lW_\ini(\loc,0) \mid \loc \in \Loc}$.
To access the thread of an event $e$, we write
$\lTID(e)$.
We let $\sR \eqdef \set{ \evt \mid \lTYP(\evt) \in \set{\lR, \lRMW}}$, and $\sW \eqdef \set{ \evt \mid \lTYP(\evt) \in \set{\lW, \lRMW}}$,
be the set of read and write events, respectively.

\inlineheadingbf{Notation} Within execution graphs, events get ordered via various relations.
For relations $R, R_1, R_2$ we write
$\inv{R}$ for the inverse relation of $R$
and $\seqR{R_1}{R_2}$ for the relational composition of $R_1$ and $R_2$.
We write $\transC{R}, \reftransC{R}$ and $\transred{R}$
for the transitive and the reflexive-transitive closure and the transitive reduction of $R$, respectively.
For a set $S$, let
$\withthread{S}{\tid} \eqdef \set{e \in S \mid \lTID(e) = \tid}$
restrict $S$ to thread~$\tid$,
and $\withthread{R}{\tid} \eqdef R \cap (\withthread{\Events}{\tid} \times \withthread{\Events}{\tid})$
restrict $R$ to pairs of events with thread~$\tid$.
Finally, $R \rst{S} \eqdef R \cap (S \times S)$
restricts $R$ to pairs over the events of $S$.

\inlineheadingbf{Litmus tests} From a set of execution graphs, we can see the {\em behavior} of a program, i.e.,
the possible values of registers at the end of the program.
Formally, behaviors   are mappings $\beta: \Reg \to \Val$ and are specified via assertions (logical formulae) over registers and values.
For such a logical formula $\varphi$, we write $\mathcal{B}_\varphi := \set{ \beta \mid \beta \models \varphi}$ for the set of
behaviors described by $\varphi$. A litmus test in the tool \herd consists of two parts: a program and a behavior assertion.
We can have two sorts of behavior assertions:
{\em required} and {\em forbidden} behaviors.\footnote{
	For the reader familiar with \herd,
	required means we require that \herd reports
	that the final state assertion holds \texttt{sometimes} or \texttt{always},
	forbidden that it holds \texttt{never}.
}
A set of behaviors $\mathcal{B}$ is {\em consistent} with a set of required and forbidden behaviors $\varphi$ ($\mathcal{B} \models \varphi$) if all of the required and none of
the forbidden behaviors are in $\mathcal{B}$.

\section{Symbolic Execution Graphs}\label{sec:symbolic-execution-graphs}

Executions of programs are modelled via (symbolic or concrete) {\em execution graphs} with events as nodes and edges describing relations.
Common to all JMMs are the relations
program order $\po$ (order of steps within one thread as per the program) and reads-from $\rf$ (to see which writes have written the values read by reads).
Typically, memory models require $\po$ and $\rf$ together to be acyclic, as to forbid out-of-thin-air-behavior.
The Java memory model, however, allows $\po \cup \rf$ to contain cycles.
As an illustration of this fact (and the way we describe such executions within symbolic execution graphs), consider the
following program \LBx (taken from~\cite[Fig.~1.3]{manson-thesis}).

\noindent\hspace*{\fill}
\begin{minipage}{.50\textwidth}
	\centering
	\begin{tabular}[b]{@{}c@{}}
		$\cloc{x}=\cloc{y}=0$ \\
		$\begin{array}{@{}l@{~}||@{~}l@{}}
			\begin{array}[t]{l}
				\readInst{\creg{a}}{\cloc{x}}{\lPln}; \\
				\assignInst{\creg{b}}{\creg{a} \ |\  1}; \\
				\writeInst{\cloc{y}}{\creg{b}}{\lPln}
			\end{array}
			&
			\begin{array}[t]{l}
				\readInst{\creg{c}}{\cloc{y}}{\lPln};
				\\
				\writeInst{\cloc{x}}{\creg{c}}{\lPln}
			\end{array}
		\end{array}$
	\end{tabular}
\end{minipage}
\hspace*{\fill}

\begin{figure}[t]
	\centering
	\begin{subfigure}{.305\textwidth}
		\centering
		\scalebox{0.9}{
			\begin{tikzpicture}[x=0.7cm]
				\node (a1) at (1.5,5) {$\lW_\ini(\cloc{x},0)$};
				\node (a2) at (3.5,5) {$\lW_\ini(\cloc{y},0)$};
				\node (b) at (1,4) {$\lR^\lPln_1(\cloc{x},\creg{a})$};
				\node (c) at (1,3) {$\lW^\lPln_1(\cloc{y},\creg{b})$};
				\node (d) at (4,4) {$\lR^\lPln_2(\cloc{y},\creg{c})$};
				\node (e) at (4,3) {$\lW^\lPln_2(\cloc{x},\creg{c})$};
				\path
				(b) edge[po] node[left] {$\po$} (c)
				(d) edge[po] node[right] {$\po$} (e)
				;

				\node[align=center] (t) at (2.5,2.21) {
					$\Constraints = \left(\creg{b} = \creg{a} \ |\ 1\right)$\\
				};
			\end{tikzpicture}
		}
		\caption{Control-flow}\label{fig:ctc8-singlet}
	\end{subfigure}
	\begin{subfigure}{.305\textwidth}
		\centering
		\scalebox{0.9}{
			\begin{tikzpicture}[x=0.7cm]
				\node (a1) at (1.5,5) {$\lW_\ini(\cloc{x},0)$};
				\node (a2) at (3.5,5) {$\lW_\ini(\cloc{y},0)$};
				\node (b) at (1,4) {$\lR^\lPln_1(\cloc{x},\creg{a})$};
				\node (c) at (1,3) {$\lW^\lPln_1(\cloc{y},\creg{b})$};
				\node (d) at (4,4) {$\lR^\lPln_2(\cloc{y},\creg{c})$};
				\node (e) at (4,3) {$\lW^\lPln_2(\cloc{x},\creg{c})$};
				\path
				(b) edge[po] node[left] {$\po$} (c)
				(d) edge[po] node[right] {$\po$} (e)
				(c) edge[rf] node[above] {$\rf$} (d)
				(e) edge[rf] node[below] {$\rf$} (b);

				\node[align=center] (t) at (2.5,2.21) {
					$\Constraints = \left(\creg{b} = \creg{a} \ |\ 1\right)$\\
					$\land\, \left(\creg{a} = \creg{c}\right) \land \left(\creg{c} = \creg{b}\right)$
				};
			\end{tikzpicture}
		}
		\caption{Data-flow}\label{fig:ctc8-bare}
	\end{subfigure}
	\begin{subfigure}{.37\textwidth}
		\centering
		\scalebox{0.9}{
			\begin{tikzpicture}
				\node (a1) at (1.7,5) {$\lW_\ini(\cloc{x},0)$};
				\node (a2) at (3.3,5) {$\lW_\ini(\cloc{y},0)$};
				\node (b) at (1,4) {$\lR^\lPln_1(\cloc{x},\creg{a})$};
				\node (c) at (1,3) {$\lW^\lPln_1(\cloc{y},\creg{b})$};
				\node (d) at (4,4) {$\lR^\lPln_2(\cloc{y},\creg{c})$};
				\node (e) at (4,3) {$\lW^\lPln_2(\cloc{x},\creg{c})$};
				\path
				(a1) edge[hb] node[left] {$\sw, \hb\ $} (b)
				(a2) edge[hb] node[below,sloped] {$\sw$} node[above,sloped] {$\hb\ $} (b)
				(a1) edge[hb] node[below,sloped] {$\sw$} node[above,sloped] {$\ \hb$}(d)
				(a2) edge[hb] node[right] {$\ \sw, \hb$} (d)
				(b) edge[po] node[left] {$\po$} node[right]{$\hb$} (c)
				(d) edge[po] node[right] {$\po$} node[left]{$\hb$} (e)
				(c) edge[rf] node[above] {$\rf$} (d)
				(e) edge[rf] node[below] {$\rf$} (b);

				\node[align=center] (t) at (2.5,2.21) {
					$\Constraints = \left(\creg{b} = \creg{a} \ |\ 1\right)$\\
					$\land\, \left(\creg{a} = \creg{c}\right) \land \left(\creg{c} = \creg{b}\right)$
				};
			\end{tikzpicture}
		}
		\caption{with \jls relations}\label{fig:ctc8-complete}
	\end{subfigure}
	\caption{
		An example of stages~1--3 of the algorithm,
		constructing the symbolic execution graph for program \LBx.
		Single thread control flow semantics~\subref{fig:ctc8-singlet},
		one symbolic execution graph~{(\subref{fig:ctc8-bare})},
		and its enhancement~{(\subref{fig:ctc8-complete})} with relations of \jls.
	}\label{fig:exec-graph}
\end{figure}

Here, all access modes are plain ($\lPln$), and $|$ is the bitwise logical ``or'' operator.
The behavior $\creg{a} = 1 \land \creg{b} = 1 \land \creg{c} = 1$ is allowed by the Java language specification~\cite{jmm-jls} 
for the  following reason: Interthread analysis could determine
that $\cloc{x}$ and $\cloc{y}$ are always either 0 or 1,
and thus determine that $\creg{b}$ is always 1.
If the constant 1 is substituted for $\creg{b}$,
then this example becomes a classic load buffering example,
where the behavior must be allowed
to be efficiently implementable on e.g.\ ARM processors.
\Cref{fig:ctc8-bare} describes the symbolic execution graph out of which we can derive such a behavior.
In this graph, we see symbolic expressions (register names) instead of concrete values being read and written.
Attached to this graph, we have a {\em constraint}, a logical expression over symbolic expressions detailing dependencies. For the graph in \Cref{fig:ctc8-bare}, the constraint would be
\[ \Constraints = \underbrace{\creg{b} = \creg{a} \ |\ 1}_{\text{assign. constraint}} \ \ \land \underbrace{\creg{a} = \creg{c}}_{\text{$\rf$-constraint}} \land \underbrace{\creg{c} = \creg{b}}_{\text{$\rf$-constraint}} \]
Note that we can use register names in symbolic expressions because of the static single assignment form of programs.

\begin{definition}
	A \emph{symbolic execution graph} $G = (E,\po,\rf,\Constraints)$ of a program $P$ has the following components:
	\begin{itemize}
		\item $E \subseteq \Events$ is a set of events containing exactly one initial write $\lW_\ini(x,0)$ for every location $x$ used in $P$.
		\item $\po \subseteq E \times E$ is the program order totally ordering the events within a thread, i.e., $\forall \tid \in \TIDs : \withthread{\po}{\tid}$
			is a total order on $\withthread{E}{\tid}$.
		\item $\rf \subseteq \sR \times \sW$ is the reads-from relation satisfying $\lLOC(e) = \lLOC(e')$ for all $(e,e') \in \rf$. Each read reads-from exactly one write: $\rf^{-1}$ is a function.
		\item $\Constraints$ is a set (or conjunction) of constraints over symbolic expressions such that $(\lSXPR(e')=\lSXPW(e)) \in \Constraints$ for all $(e,e') \in \rf$.
	\end{itemize}
\end{definition}

\noindent For a graph $G$, we write $G.E$ for its events, $G.W$ for its write events, $G.\po$ for its program order and so on.
A concrete execution graph is a symbolic execution graph
where each expression is a constant
and each read is justified by a write with the same value,
that is, $\forall (e,e') \in \rf : \lSXPW(e) =\lSXPR(e') \in \Val$.
We obtain a concrete execution graph from a symbolic execution graph
by finding a mapping $\gamma: \Reg \rightarrow \Val$ which is a logical model of $\Constraints$ and then
substituting all symbolic expressions in events by their instantiations via $\gamma$.

\paragraph{Building execution graphs.}
Similar to \herd, we build execution graphs in multiple stages.
However, while \herd attempts to build execution graphs with concrete values,
and fails to properly evaluate assertions
for graphs where no unique value can be found by forward substitution,
we use symbolic execution graphs to handle these cases,
as they are highly relevant to MEMs, specifically JMMs.

\inlineheadingbf{1.\ Control-flow semantics}
First, we construct graphs per thread by following
the thread's program according to a sequential semantics.
Initially, the thread graph is empty, i.e., $G_0 = (\emptyset,\emptyset,\emptyset,\emptyset)$.
Given a graph $G = (E,\po,\rf,\Constraints)$ representing the execution so far, we
extend it depending on the type of the next instruction of thread $\tid$.
All newly added events are appended $\po$-after the previous events. Relation $\rf$ remains empty in this stage.

\begin{description}
	\item[Assignment] For an assignment of the form $\assignInst{\reg}{\expr}$,
		we extend $\Constraints$ by the constraint $(\reg=\expr)$. No additional events are added.
	\item[Store]
		For a store $\writeInst{\loc}{\sexpr}{\textsf{wm}}$,
		we add an event labeled $\wlab{\wmd}{\loc}{\sexpr}$
		with thread id $\tid$
		and a fresh identifier,
		without adding constraints.
	\item[Load]
		For a load $\readInst{\reg}{\loc}{\textsf{rm}}$,
		we add an event labeled $\rlab{\rmd}{\loc}{\reg}$,
		with thread id $\tid$
		and -- as for all events -- a fresh identifier,
		without adding constraints.
	\item[RMW] For a compare-and-exchange instruction $\casInst{\reg}{\loc}{\sexpr_1}{\sexpr_2}{\textsf{rm,wm}}$,
		we have to construct two possible extensions, one where the CAX succeeds and the other
		where it fails. In the first case, we add an event labelled $\rmwlab{\rmd,\wmd}{\loc}{\sexpr_1}{\sexpr_2}$
		and a constraint of the form $(\sexpr_1 = \sexpr_2)$,
		in the second case an event labelled $\rlab{\rmd}{\loc}{\sexpr_1}$
		and a constraint $(\sexpr_1 \neq \sexpr_2)$.
	\item[Fence] For a fence $\fenceInst{\textsf{fm}}$, we add an event labelled $\flab{\fmd}$ and no constraints.
	\item[If-then-else] Like for RMW, we build two successor graphs, one with additional constraint $\mathit{cond}$ (being the condition of the if) and one with $\neg \mathit{cond}$.
\end{description}

\noindent The procedure ends once no instructions remain.
Since the focus of weak memory models is on values read from and written to shared locations,
termination of the procedure requires the absence of loops whose behavior depends on memory accesses.
Consequently, our tool,
like most other weak-memory tools,
supports only loop-free programs.
At that point, we have all conceivable executions of a thread, i.e., for every thread $\tid$ we have a set of graphs $\mathcal{G}_\tid$.
The program executions of the parallel composition of the threads then correspond
to the Cartesian product of the graphs of all threads.
More precisely, for each family of graphs $\tup{G_\tid = (E_\tid, \po_\tid, \emptyset,\Constraints_\tid)}_{\tid \in \TIDs}$, one per thread, we
construct a new product graph $G = (\bigcup_{\tid \in \TIDs} E_ \tid, \bigcup_{\tid \in \TIDs}
\po_\tid, \emptyset, \bigcup_{\tid \in \TIDs} \Constraints_\tid)$. To this graph,
we finally add initializing events $\lW_\ini(\loc,0)$ for every $\loc \in \Loc$.

An example of the result of stage~1 is shown in \Cref{fig:ctc8-singlet}.
Since the \LBx program has no branching and no compare-and-exchange instructions,
this is the only result.

\inlineheadingbf{2.\ Data-flow semantics}
In stage~2, we add the $\rf$ edges.
Out of every graph $G = (E,\po,\rf,\Constraints)$ which stage~1 produces, we construct several graphs by adding all possible $\rf$ relations.
More specifically, for every $r \in E \cap \sR$, we choose some $w \in E \cap \sW$ such that
$\lLOC(r) = \lLOC(w)$, add $(w,r)$ to $\rf$ plus add a constraint $(\lSXPR(r)=\lSXPW(w))$ to $\Constraints$.
Whenever the resulting set of constraints $\Constraints$ is unsatisfiable (which is checked using an SMT solver), we entirely remove the graph from our set of considered graphs. Note that we have no further requirements on $\po$ and $\rf$ here; this is what a concrete
single-execution model adds.

One of the graphs created by stage~2,
where each read reads from a non-initial write,
is shown in \Cref{fig:ctc8-bare}.
As the result of stage~2, we also get 3 other graphs,
one where all read events read from initial writes,
and two where one of the read events read from an initial and the other from a non-initial write.

\inlineheadingbf{3.\ Single execution memory model}
So far, the constructed graphs are independent of any SEM.
In the last step, we now take the set of graphs $\mathcal{G}$ computed as the result of stage~2, a SEM in the form of a \cat model of \herd and
remove all graphs inconsistent with the requirements of the SEM.

Typically, memory models define more relations than just $\po$ and $\rf$. With a SEM fixed, we thus build {\em enhanced} symbolic execution graphs.
As an example, we use \jls (the complete \cat model can be found in
\ifthenelse{\boolean{hideAppendix}}{the extended version of this paper}{\Cref{sec:appendix-JMM}}).
\jls defines a relation $\hb$ (happens-before) by
\verb+let hb = ( po | sw )*+,
i.e., defines $\hb = (\po \cup \sw)^*$, where $\sw$ is the relation synchronizes-with.
It then states the consistency requirement \verb+irreflexive rf;hb+.
A number of other relations and constraints fix further requirements.

\Cref{fig:ctc8-complete} shows the enhancement of the graph of \Cref{fig:ctc8-bare} with relations of $\jls$.
Note that -- due to the symbolic expressions -- this graph and its constraint represent an infinite number of concrete graphs,
namely all those where $a=b=c=\val$ for some odd value $\val \in \Val$.

\section{Checking Causality Requirements}\label{sec:causality-requirements}

Next, we build the multi-execution model out of the set of graphs $\Executions$ constructed according to a SEM.
The overall objective of the construction is to determine which graphs are {\em valid}, thereby ruling out unwanted out-of-thin-air behavior.
As an example, consider the graph in \Cref{fig:ctc8-complete} of program \LBx.
This graph has a potentially problematic {\em race} (actually two): Event $\lW^\lPln_2(\cloc{x},\creg{c})$ has an $\rf$ edge to event $\lR^\lPln_1(\cloc{x},\creg{a})$
and these two events are not ordered via relation happens-before $\hb$.
The question to be answered is whether a JMM considers this to be ok or not. To this end,
the Java specification has introduced the concept of {\em causality checking}, which we
here extend to symbolic execution graphs.

Our tool expects the SEM to have constructed {\em well-formed} graphs $G$ for a program containing
program order $\po$ and reads-from relation $\rf$ plus relations $\so$ (synchronization order), $\sw$ (synchronizes-with)
and $\hb$ (happens-before), i.e., $G = (E, \po, \so, \rf, \sw, \hb, \Constraints)$, and the additional relations have to satisfy

\begin{itemize}
	\item $\hb$ is a partial order, $\sw \subseteq \hb$ and
		$\so$ totally orders the volatile events.
\end{itemize}

\noindent The multi-execution part of the memory model now considers multiple execution graphs
to determine the validity of a single graph.
To this end, it has to relate events of {\em different} execution graphs.
For this, we define an equivalence relation on events of (potentially different) graphs\footnote{To our knowledge, no existing JMM literature formally defines such an equivalence.}.
First, we consider two events $e_1 \in E_1, e_2 \in E_2$ to be \emph{similar},
denoted $e_1 \sim e_2$,
if $\lTYP(e_1) = \lTYP(e_2)$
and -- if they operate on a location --
$\lLOC(e_1) = \lLOC(e_2)$.
For each event, we count the number of similar events preceding it in program order within $G$,
giving its {\em index}: $\mathrm{idx}(e,G)$.
If $e \in G.E$ is an initial write action,
then we define $\mathrm{idx}(e,G) = 0$.
Otherwise, $e$ has an executing thread $\tid \in \TIDs$
and $G.\po \rst{\tau}$ is a total order.
We define $\mathrm{idx}(e,G)$ to be the position of $e$ in $G.\po \rst{\tid}\; \cap \sim$.
Finally, events $e_1$ and $e_2$ of graphs $G_1$ and $G_2$ are {\em equal}
if they are similar and have the same position:
$e_1 = e_2 \iff {e_1 \sim e_2} \land \mathrm{idx}(e_1,G_1) = \mathrm{idx}(e_2,G_2)$.
In the sequel all operations on elements, sets and relations
are performed {\em modulo this equivalence relation}.

The MEM now takes the set of well-formed execution graphs $\Executions$
and rules out those that are considered unreasonable.
Let $T \in \Executions$ be a {\em target} which we intend to check for validity.
Causality checking incrementally constructs a {\em justification} for $T$ from our set of graphs $\Executions$:
we start the construction with a well-behaved execution $G \in \Executions$ in which all $\rf$-related events are also
in happens-before relation: $\forall (w,r) \in G.\rf: (w,r) \in G.\hb$.
Then, we incrementally {\em commit} events. Committing basically fixes values
written or read by events (in our case, we fix symbolic values).
The Java language specification defines a number of conditions on commitments which we have implemented in \ToolName\footnote{If a user of \ToolName wishes to use a different set of conditions, the code must be edited.
For conditions such as those of~\cite{tphol/AS07} this is straightforward.}
The check for validity of $T$ ends with success when we have managed to commit all of $T$'s events.
More formally, a target $T$ is allowed by the memory model,
if there is a so called {\em justification sequence} $G_1, \dots, G_n$
where each $G_i$ is well-formed,
together with a sequence of committed events $\Committed_0 \subseteq \ldots \subseteq \Committed_n \subseteq \Events $
where $\Committed_0 = \emptyset$, and $\Committed_n$ = $G_n.E$.
In addition, the following six conditions (of \jls) need to hold for every $i$,
which we capture in a predicate
$\mathsf{ok}(T,\tup{G_i,\Committed_i},\Committed_{i-1})$:

\begin{enumerate}
	\item[1.] $\Committed_i \subseteq G_i.E$ (committed events occur in the graph),
	\item[2.] $G_i.\hb \rst{\Committed_i} = T.\hb \rst{\Committed_i}$ ($\hb$ over committed events $C_i$ is the same in $G_i$ and $T$),
	\item[3.] $G_i.\so \rst{\Committed_i} = T.\so \rst{\Committed_i}$ ($\so$ over committed events $C_i$ is the same in $G_i$ and $T$),
	\item[5.] $G_i.\rf \rst{\Committed_{i-1}} = T.\rf \rst{\Committed_{i-1}}$ ($\rf$ over committed events $C_{i-1}$ agrees in $G_i$ and $T$),
	\item[6.] $\forall r \in G_i.R \setminus \Committed_{i-1} :
		(G_i.\invrf(r), r) \in G_i.\hb$ (reads not committed in stage $i-1$ must be in $\hb$-order with their writes),
	\item[7.] $\forall r \in (\Committed_i \setminus \Committed_{i-1}) \cap \sR :
		G_i.\invrf(r) \in \Committed_{i-1} \land T.\invrf(r) \in \Committed_{i-1}$ (reads can only be committed after their writes).
\end{enumerate}

\noindent The \jls additionally has conditions numbered 4, 8 and 9.
Condition 9 is about external operations (like print statements)
which, for simplicity, we ignore here;
Condition 8 is given below and only checked after justifications are generated.
Condition 4 requires that $V_i \rst{\Committed_i} = V \rst{\Committed_i}$ for all $i$,
where $V$ is a function mapping each write $w$ to the written value.
Basically, condition~4 is saying that once we have committed an event,
its value may not change anymore.
Since we have symbolic execution graphs, we might not only have values, but also symbolic expressions. Our tool thus first ignores condition 4 during construction of justification sequences and then 
checks it in a final validation step (see below).

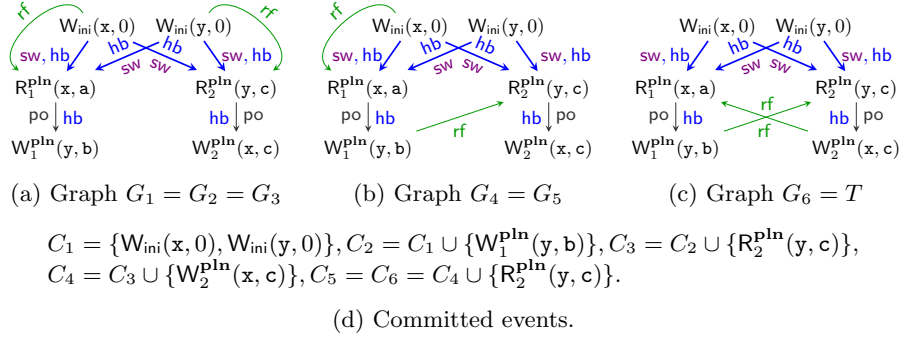
\begin{figure}[t]
	\begin{subfigure}[b]{.33\textwidth}
		\centering
		\scalebox{0.8}{
			\begin{tikzpicture}[bezier bounding box]
				\node (a1) at (1.7,5) {$\lW_\ini(\cloc{x},0)$};
				\node (a2) at (3.3,5) {$\lW_\ini(\cloc{y},0)$};
				\node (b) at (1,4) {$\lR^\lPln_1(\cloc{x},\creg{a})$};
				\node (c) at (1,3) {$\lW^\lPln_1(\cloc{y},\creg{b})$};
				\node (d) at (4,4) {$\lR^\lPln_2(\cloc{y},\creg{c})$};
				\node (e) at (4,3) {$\lW^\lPln_2(\cloc{x},\creg{c})$};
				\path
				(a1) edge[hb] node[left] {$\sw, \hb$} (b)
				(a2) edge[hb] node[below,sloped] {$\sw$} node[above,sloped] {$\hb\ $} (b)
				(a1) edge[hb] node[below,sloped] {$\sw$} node[above,sloped] {$\ \hb$}(d)
				(a2) edge[hb] node[right] {$\sw, \hb$} (d)
				(b) edge[po] node[left] {$\po$} node[right]{$\hb$} (c)
				(d) edge[po] node[right] {$\po$} node[left]{$\hb$} (e)
				(a1) edge[rf, bend angle=100, bend right] node[above, sloped] {$\rf$} (b)
				(a2) edge[rf, bend angle=100, bend left] node[above,sloped] {$\rf$} (d);
			\end{tikzpicture}
		}
		\caption{Graph $G_1 = G_2 = G_3$}
	\end{subfigure} \hfill
	\begin{subfigure}[b]{.33\textwidth}
		\centering
		\scalebox{0.8}{
			\begin{tikzpicture}[bezier bounding box]
				\node (a1) at (1.7,5) {$\lW_\ini(\cloc{x},0)$};
				\node (a2) at (3.3,5) {$\lW_\ini(\cloc{y},0)$};
				\node (b) at (1,4) {$\lR^\lPln_1(\cloc{x},\creg{a})$};
				\node (c) at (1,3) {$\lW^\lPln_1(\cloc{y},\creg{b})$};
				\node (d) at (4,4) {$\lR^\lPln_2(\cloc{y},\creg{c})$};
				\node (e) at (4,3) {$\lW^\lPln_2(\cloc{x},\creg{c})$};
				\path
				(a1) edge[hb] node[left] {$\sw, \hb$} (b)
				(a2) edge[hb] node[below,sloped] {$\sw$} node[above,sloped] {$\hb\ $} (b)
				(a1) edge[hb] node[below,sloped] {$\sw$} node[above,sloped] {$\ \hb$}(d)
				(a2) edge[hb] node[right] {$\ \sw, \hb$} (d)
				(b) edge[po] node[left] {$\po$} node[right]{$\hb$} (c)
				(d) edge[po] node[right] {$\po$} node[left]{$\hb$} (e)
				(a1) edge[rf, bend angle=100, bend right] node[above, sloped] {$\rf$} (b)
				(c) edge[rf] node[below] {$\rf$} (d);
			\end{tikzpicture}
		}
		\caption{Graph $G_4 = G_5$}
	\end{subfigure} \hfill
	\begin{subfigure}[b]{.32\textwidth}
		\centering
		\scalebox{0.8}{
			\begin{tikzpicture}
				\node (a1) at (1.7,5) {$\lW_\ini(\cloc{x},0)$};
				\node (a2) at (3.3,5) {$\lW_\ini(\cloc{y},0)$};
				\node (b) at (1,4) {$\lR^\lPln_1(\cloc{x},\creg{a})$};
				\node (c) at (1,3) {$\lW^\lPln_1(\cloc{y},\creg{b})$};
				\node (d) at (4,4) {$\lR^\lPln_2(\cloc{y},\creg{c})$};
				\node (e) at (4,3) {$\lW^\lPln_2(\cloc{x},\creg{c})$};
				\path
				(a1) edge[hb] node[left] {$\sw, \hb$} (b)
				(a2) edge[hb] node[below,sloped] {$\sw$} node[above,sloped] {$\hb\ $} (b)
				(a1) edge[hb] node[below,sloped] {$\sw$} node[above,sloped] {$\ \hb$}(d)
				(a2) edge[hb] node[right] {$\sw, \hb$} (d)
				(b) edge[po] node[left] {$\po$} node[right]{$\hb$} (c)
				(d) edge[po] node[right] {$\po$} node[left]{$\hb$} (e)
				(c) edge[rf] node[above] {$\rf$} (d)
				(e) edge[rf] node[below] {$\rf$} (b);
			\end{tikzpicture}
		}
		\caption{Graph $G_6=T$}
	\end{subfigure}
\begin{subfigure}{\textwidth}
	\centering
	\medskip
	$\begin{array}{l}	
		C_1 = \{\lW_\ini(\cloc{x},0),\lW_\ini(\cloc{y},0)\},
		C_2 = C_1 \cup \{\lW^\lPln_1(\cloc{y},\creg{b})\},
		C_3 = C_2 \cup \{\lR^\lPln_2(\cloc{y},\creg{c})\},\\
		C_4 = C_3 \cup \{\lW^\lPln_2(\cloc{x},\creg{c})\},
		C_5 = C_6 = C_4 \cup \{\lR^\lPln_2(\cloc{y},\creg{c})\}.
	\end{array}$
	\caption{Committed events.}
	\end{subfigure}
	\caption{Justification sequence for the execution graph $T$ of \Cref{fig:ctc8-complete}.}
	\label{fig:justification}
\end{figure}

\inlineheadingbf{4.\ Generating Justifications}
Stage 4 in the construction of valid execution graphs out of programs is now
the actual generation of justification sequences.
Formally, a justification sequence is a sequence of \emph{justification stages}
which are pairs
$s = \tup{G,\Committed}$
where $G$ is an execution graph
and $\Committed \subseteq \Events$ is a set of committed events.
A \emph{justification} for a target execution graph $T$ is a pair
$J = \tup{T,(s_1,\dots,s_n)}$
where $(s_1,\dots,s_n)$ is a justification sequence.

For a given target graph $T$,
a current justification stage $\tup{G,\Committed}$
and a collection of well-formed execution graphs $\Executions$,
we compute the successor justification stages with
$\mathrm{GenSuccessors}(T,\tup{G,\Committed},\Executions) =$
\[
\bigcup_{G' \in \Executions}
\bigcup_{\emptyset \neq X \subseteq G'.E \setminus \Committed}
\set{\tup{G',\Committed'}
\mid \Committed' = (\Committed \cap G'.E) \cup X
\land \mathsf{ok}(T,\tup{G',\Committed'},\Committed)
}.
\]

\noindent We always start a justification sequence with a \emph{well-behaved} execution $G_1 \in \Executions$.
We set $C_0$ (which never gets constructed) to $\emptyset$.
For graph $G_1$,
the initial stage is
$s_1 = \tup{G_1, G_1.E \rst \ini}$, the stage with just the initial writes committed.
We then generate the justification sequences by exploring all finite sequences $(s_1,\dots,s_n)$
such that for all $1 < i < n$,
$s_{i+1} \in \mathrm{GenSuccessors}(T,s_i,\Executions)$.
A justification sequence $(s_1,\dots,\tup{G_k,\Committed_k})$
generated in this way is {\em valid} (and the target $T$ is valid) iff
$\lvert \Committed_k \rvert = \lvert T.E \rvert$ and
$\mathsf{ok}(T,\tup{T,T.E},\Committed_k)$ and condition~8 is fulfilled: 
\begin{itemize}
	\item[8.] if
		$(x,y) \in G.\sw \cap (\transred{G_i.\hb} \setminus G_i.\po)
		\land (y,z) \in G_i.\hb
		\land z \in \Committed_i$,
		then $(x,y) \in G_j.\sw$ for all $j \geq i$.
\end{itemize}
Condition~8 enforces the following constraint: 
once an action $z$ is committed,
 removing  the release-acquire synchronization
that was necessary to justify why $z$ is allowed to see what it sees is not possible anymore. 

As an example of stage 4, consider program \LBx and one of its symbolic execution graphs $T$ as depicted in \Cref{fig:ctc8-complete}.
\Cref{fig:justification} shows the graphs of a justification sequence for $T$ with corresponding committed events sets.
As seen, events are incrementally committed until finally $C_6 = T.E$ and $G_6 = T$.

\inlineheadingbf{5.\ Validating Symbolic Justifications}
Next, we need to take condition 4 into account which
we carry out together with checking for specific behaviors.
Our check has to guarantee that (1) there is some instantiation of the
registers occurring in symbolic expressions with values such
that values of committed events do not change anymore over a justification sequence,
(2) the target $T$ allows for the behavior assertion, and (3) the constraints of $T$, $T.\Constraints$, are satisfied.
The check proceeds by encoding these three properties into a large first-order logic formula and
asking an SMT solver for a model of the formula.
We encode each read and write of the graphs into a variable of the form
$\mathit{graph\_name}{:}\mathit{thread}{:}\mathit{type}\,\mathit{location}@\mathit{index}$,
for example, $G_2{:}T_3{:}\rdx{\cloc{y}}@2$
is the variable representing the 2nd read of location~$\cloc{y}$
performed by thread~$T_3$ in execution graph~$G_2$.
We encode the entire justification sequence and target $T$
in such a way that the SMT solver finds a satisfying assignment to these variables,
whenever an instantiation of symbolic register names in the justification sequence with the values of the corresponding variables yields a concrete justification
 demonstrating that the behavior as specified in the assertion is possible.

The encoding works as follows. For a register~$a$ occurring as symbolic value in a graph $G$, we define a local variable
$\mathrm{ELV}_G(a) \eqdef G{:}a$.
We denote by $\phi[\mathrm{ELV}_G]$ the replacement of each register in formula $\phi$
by $\mathrm{ELV}_G$.
We encode a non-initializing\footnote{
	Initializing writes write a constant
	that is the same for all executions of a program,
	so condition 4 cannot be violated.
}
write $w \in G_i.\Writes \setminus G_i.E \rst \ini$ to location $\cloc{x}$
in justification stage $s_i = \tup{G_i,\Committed_i}$
by thread $\tid$
to location $\cloc{x}$
with write expression $\mWexpr$
and $\mathrm{idx}(w,G_i) = p$
as
\[
	\mathrm{EncWr}(w,s_i) \eqdef
	\begin{cases}
		G_i{:}\tid{:}\wrx{\cloc{x}}@p
		= \mWexpr[\mathrm{ELV}_{G_i}]
		= T{:}\tid{:}\wrx{\cloc{x}}@p
		& \text{if } w \in \Committed_i \\
		G_i{:}\tid{:}\wrx{\cloc{x}}@p
		= \mWexpr[\mathrm{ELV}_{G_i}]
		& \text{otherwise.}
	\end{cases}
\]
We encode a read $r \in G_i.\Reads$
in justification stage $s_i = \tup{G_i,\Committed_i}$
by thread $\tid$
from location $\cloc{x}$
with read expression $\mRexpr$
and $\mathrm{idx}(r) = p$
as $\mathrm{EncRd}(r,s_i) \eqdef
G_i{:}\tid{:}\rdx{\cloc{x}}@p = \mRexpr[\mathrm{ELV}_{G_i}]$.

We take the conjunction
of the encoding of all reads and non-initializing writes
in a justification stage,
and the graphs constraints $\Constraints$,
to encode the stage
$\mathrm{EncStage}(\tup{G_i,\Committed_i}) \eqdef$
\[
	\bigwedge_{w \in G_i.\Writes \setminus G_i.E\rst{\ini}}
	\mathrm{EncWr}(e,\tup{G_i,\Committed_i})
	\land \bigwedge_{r \in G_i.\Reads}
	\mathrm{EncRd}(e,\tup{G_i,\Committed_i})
	\land G_i.\Constraints[\mathrm{ELV}_{G_i}].
\]
We take the conjunction of this encoding for all justification stages of the justification
and the target stage.
We also take the conjunction with the behavior assertion $\phi$ substituted with $\mathrm{ELV}_T$,
to ensure that the justification works for a concrete target execution
satisfying the assertion.
This encodes the entire justification
$\mathrm{EncJust}(\tup{T,(s_1,\dots,s_n)},\phi) \eqdef$
\[
	\phi[\mathrm{ELV}_T]
	\land \mathrm{EncStage}(\tup{T,T.E})
	\land \bigwedge_{1 \leq i \leq n} \mathrm{EncStage}(s_i).
\]
This formula might have a satisfying assignment
mapping each read and write expression to a value.
If so, substituting these values into the justification gives a concrete justification
satisfying all conditions, including condition 4,
showing that a concrete execution satisfying the assertion is acceptable.

For our example program \LBx, we can thus see that the behavior $a=1 \wedge b=1 \wedge c=1$ is possible.
Note that an assignment with $b \mapsto 5$ is for instance not possible: in the justification sequence of
\Cref{fig:justification}, graph $G_2$ has the (non shown) constraint $b=1$.
Since $\lW_1^\lPln(\cloc{y},\creg{b}) \in C_2$ (i.e., committed), this value has to be
kept over the entire justification sequence.

 With the construction of symbolic execution graphs and causality checking,
we now have all the tooling at hand for application scenario \circled{1}.
Next, we investigate application scenario \circled{2}.

\section{Compilation Scheme Correctness} \label{sec:comp}
The second key objective of our work
is to validate that a candidate Java memory model
is compatible with a realistic compilation scheme
and associated hardware memory model.
Intuitively, soundness requires no compiled program to exhibit behavior forbidden by its source.
Formally, a compilation scheme~$\mathit{Comp}$
from a source Java memory model~$\mathit{SEM}$ (and its multi-execution version $\mathit{MEM(SEM)}$)
to a target hardware model~$H$ is sound if,
for every source program~$P$,
\begin{align}
	\mathrm{Beh}_H(\mathit{Comp}(P)) \subseteq \mathrm{Beh}_{\mathit{MEM(SEM)}}(P), \label{eq:sound}
\end{align}
where $\mathrm{Beh}_{M}(P)$
denotes the set of behaviors of $P$ under model $M$.
Our notion of compilation correctness coincides with the intent of~\cite{jam21}:
``A correct compilation, intuitively, should not introduce any new program behavior,''
although their formalization is different from ours.
{\em Proving} this inclusion is expensive and error-prone
(as seen in \Cref{sec:evaluation}).
To support fast, iterative development of models,
\ToolName offers an automated \emph{testing-based} approximation.
It can expose unsoundness by finding a counterexample, 
i.e., a behavior of $\mathit{Comp}(P)$ not permitted by $\mathit{MEM(SEM)}$,
but cannot prove soundness.

\ToolName takes as input a source $\mathit{SEM}$ in \herd's \cat format
and a program $P$ written in \herd's litmus test format for Java.\footnote{
	Although the behavior assertion is not of interest in scenario \circled{2},
	we keep this format
	because that way we can use the same input file for all use-cases of \ToolName.
}
The tool outputs either a \emph{pass} (no counterexamples)
or a \emph{fail} verdict, with a witness behavior violating the inclusion.
We currently target the x86 architecture due to its simplicity,
appearance in~\cite{jam21},
and strong \herd support.
Instead of using an existing compiler,
we implemented a simple,
non-optimizing compiler
to isolate the compilation scheme itself.
We briefly describe this compiler.
Each source instruction is translated into a sequence of x86 instructions
according to the scheme \ifthenelse{\boolean{hideAppendix}}{given in the extended paper}{in \Cref{sec:x86-compilation-scheme}},\footnote{
	The compilation scheme of~\cite{jam21} coincides with ours on common instructions.
}
and source registers are systematically mapped to x86 registers.
We track which thread each register belongs to,
since in x86 each thread has its own register set.
For instance,
$\creg{a}$ in thread $\ctid{1}$ may map to $\creg{EAX}$ of $\ctid{1}$,
while $\creg{b}$ in thread $\ctid{2}$ maps to $\creg{EAX}$ of $\ctid{2}$.
For registers appearing in behavior assertions,
this mapping is bijective, hence invertible.
\begin{note}
	One might notice that the behavior assertion should not matter here,
	but \herd in its listing of allowed behaviors
	only outputs the values of registers which appear in the behavior assertion.
	Therefore, we will only obtain and thus only need to invert
	mappings for registers that appear in the behavior assertion.
\end{note}

\noindent The soundness check of behavior inclusion as stated in (\ref{eq:sound}) proceeds as follows:
\begin{enumerate}
	\item \textbf{Compilation}:
		Source program $P$ is compiled into an x86 program $\mathit{Comp}(P)$.
	\item \textbf{Target hardware evaluation}:
		\herd enumerates all allowed behaviors $\mathrm{Beh}_H(\mathit{Comp}(P))$ under the x86 memory model.
	\item \textbf{Restoration}:
		The found behaviors are parsed,
		and the register mapping is inverted to express each behavior
		in terms of the original source registers.
	\item \textbf{Inclusion check}:
		For each behavior $\beta \in \mathrm{Beh}_H(\mathit{Comp}(P))$,
		\ToolName checks whether $\beta \in \mathrm{Beh}_{\mathit{MEM(SEM)}}(P)$.
		Rather than full, upfront enumeration of $\beta$,
		\ToolName queries the source model on demand,
		avoiding exploration of unreachable behaviors
		that are costly due to MEM complexity.
	\item \textbf{Verdict and witness}:
		If any behavior~$\beta$ fails the check,
		the tool reports \emph{fail} and provides $\beta$ as a witness.
		Otherwise, the test \emph{passes}.
\end{enumerate}

\section{Tool \ToolName}

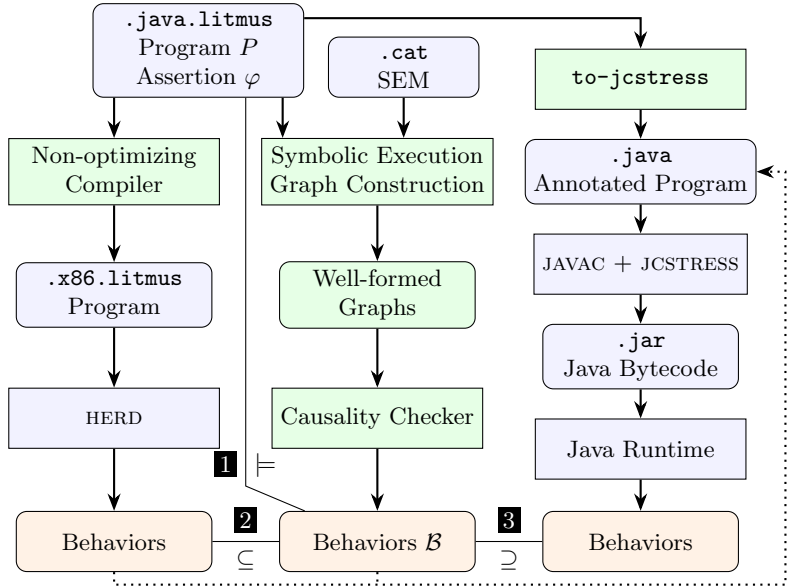
\begin{figure}[t]
	\centering
	\begin{tikzpicture}[
  x=3.5cm,y=0.082cm,
  doc/.style={
    draw,
    rectangle,
    rounded corners,
    fill=white,
    minimum width=2.6cm,
    minimum height=0.8cm,
    align=center,
    font=\small
  },
  actor/.style={
    draw,
    rectangle,
    fill=white,
    minimum width=2.8cm,
    minimum height=0.8cm,
    align=center,
    font=\small
  },
  arrow/.style={
    ->,
    thick,
    >=Stealth
  },
	sota/.style={ 
		fill=blue!5
	},
	result/.style={
		fill=orange!10
	},
	ours/.style={
		fill=green!10
	}
]
	\node[doc,sota, anchor=south, minimum width=2.8cm] (java) at (-0.68,12) {\texttt{.java.litmus}\\Program $P$\\Assertion $\varphi$};
	\node[doc,sota, anchor=south,minimum width=2cm] (cat) at (0.1,12) {\cat\\SEM};

	\node[actor,ours] (sym) at(0,-0) {Symbolic Execution\\Graph Construction};
	\node[doc,ours] at (0,-20) (graphs) {Well-formed\\Graphs};
	\node[actor,ours] (cause) at (0,-40) {Causality Checker};
	\node[doc,result] (behM) at (0,-60){Behaviors~$\mathcal{B}$};

	\node[actor,ours] (compiler) at (-1,-0) {Non-optimizing\\Compiler};
	\node[doc,sota] (x86) at(-1,-20) {\texttt{.x86.litmus}\\Program};
	\node[actor,sota] (herd) at (-1,-40) {\herd};
	\node[doc,result] (behL) at (-1,-60) {Behaviors};

	\node[actor,ours] (tojc) at (1,15) {\texttt{to-jcstress}};
	\node[doc,sota] (annot) at (1,-0) {\texttt{.java}\\Annotated Program};
	\node[actor,sota] (javac) at (1,-15) {\textsc{javac} + \jcstress};
	\node[doc,sota] (jar) at (1,-30) {\texttt{.jar}\\Java Bytecode};
	\node[actor,sota] at (1,-45) (runtime) {Java Runtime};
	\node[doc,result] at (1,-60) (behR) {Behaviors};

	\coordinate (mid) at ($ (java.east)!0.5!(sym.west) $);
	\draw[arrow] (java.south -| mid) -- (sym.north -| mid);
	\draw[arrow] (cat) -- (sym.north -| cat.south);
	\draw[arrow] (sym) -- (graphs);
	\draw[arrow] (graphs) -- (cause);
	\draw[arrow] (cause) -- (behM);

	\draw[arrow] (java.south -| compiler.north) -- (compiler.north);
	\draw[arrow] (compiler) -- (x86);
	\draw[arrow] (x86) -- (herd);
	\draw[arrow] (herd) -- (behL);

	\draw[arrow] ([yshift=-2mm] java.north east) -| (tojc);
	\draw[arrow] (tojc) -- (annot);
	\draw[arrow] (annot) -- (javac);
	\draw[arrow] (javac) -- (jar);
	\draw[arrow] (jar) -- (runtime);
	\draw[arrow] (runtime) -- (behR);

	\draw[dotted,thick] (behM) |- (0,-67);
	\draw[dotted,arrow] (behL) |- (0,-67) -| (1.55,-67) |- (annot);

	\path
		(behL) edge node[below] {$\subseteq$} node[above] {\circled{2}} (behM)
		(behR) edge node[below] {$\supseteq$} node[above] {\circled{3}} (behM)
		;
	\coordinate (helper) at (-0.5,-51);
	\draw (java.south -| helper.center) |- node[above left] {\circled{1}} node[above right] {$\models$} (helper) -- (behM);
\end{tikzpicture}
	\caption{
		Overview of \ToolName
		(components from others in {\color{blue!70!black}blue}, ours in  {\color{green!70!black}green}).
	}
	\label{fig:overview}
\end{figure}

Next, we shortly explain \ToolName's
 overall structure and usage.

\inlineheadingbf{Tool \ToolName} \Cref{fig:overview} gives an overview of the toolchains of \ToolName.
It depicts all three application scenarios. All scenarios start with a litmus test consisting of a program $P$ and some assertion $\varphi$
of required or forbidden behavior
plus some single execution Java memory model $\mathit{SEM}$.
Scenario \circled{1} uses the behavior assertion; for scenarios \circled{2} and \circled{3} it is of no importance.

In all three scenarios, \ToolName builds the set of possible behaviors $\mathcal{B}$\footnote{For optimization purposes, we actually do not construct {\em all} behaviors, but simply check inclusion element-wise.}
of $P$ under $\mathit{MEM(SEM)}$.
In scenario \circled{1}, we check whether the required and forbidden behaviors as stated in $\varphi$ are consistent with $\mathcal{B}$ (denoted $\models$).
When applicable, \ToolName reports counterexamples.

Scenario \circled{2} uses the non-optimizing compiler to x86 and the tool \herd as described in \Cref{sec:comp},
and at the end compares the behaviors generated by \herd against $\mathcal{B}$.
Again, counterexamples (of behaviors $\beta$ allowed on x86 after compilation, but not contained in $\mathcal{B}$) are reported.

Finally, scenario \circled{3} is similar to scenario \circled{2}, but now uses an (existing) optimizing compiler generating
Java bytecode and during runtime optimized machine code. Here, we cannot use \herd anymore and instead have decided to employ the tool \jcstress~\cite{jcstress}.
\jcstress is a tool designed to aid in the validation of concurrency support for Java.
Within our toolchain, we first generate such \jcstress annotated Java programs
which are then compiled into bytecode.
When executed, the program outputs a log with the observed behavior which
can then again be compared against $\mathcal{B}$.
Optionally (dotted line), our toolchain can directly feed either the behavior as generated by the
non-optimizing compiler and \herd or $\mathcal{B}$ into the generation
of the annotated program.
In that case, checking behavior inclusion is directly part of the annotated program,
making the test highly portable.

\inlineheadingbf{Usage}
\ToolName is a command line tool.
In its most basic form,
\ToolName takes as input the program given in the \herd litmus test format for Java
and a SEM in the \cat format of \herd.
It will then check whether the behavior assertion of the litmus test is satisfied (scenario \circled{1}).
Further options, e.g., to show the justification,
are explained in \ToolName's documentation.
A number of programs can be used to execute parts of
the toolchain of \Cref{fig:overview}.
\begin{itemize}
\item \texttt{to-herd-x86} generates the \texttt{.x86.litmus} programs
employed in scenario~\circled{2}.

\item \texttt{with-herd-x86} builds on the former,
but instead of displaying the generated x86 program
it gives it to \herd
to compute the allowed behaviors.

\item \texttt{to-javasim} converts the program given in the \herd litmus test format for Java
to the input format of the tool \JavaSim~\cite{ftfjp/MP02}.
We need this conversion for our evaluation in \Cref{sec:evaluation}.

\item \texttt{with-javasim}:
Similarly to \texttt{with-herd-x86},
this program uses \JavaSim to generate the allowed behaviors of the program.

\item \texttt{to-jcstress} converts the program given in the \herd litmus test format for Java
to a Java program ready for use with \jcstress.
\end{itemize}

\section{Evaluation}\label{sec:evaluation}

\begin{samepage}
Our evaluation of \ToolName is guided by the following research question:
\begin{center}
\textbf{RQ}: Is \ToolName able to detect correctness issues in Java memory models?
\end{center}
\end{samepage}

\noindent More specifically, we are interested in the tool's  usefulness within our three application scenarios.
To this end, we have chosen three Java memory models to test during the evaluation:
The official Java language specification
(\jls, \cite{jmm-jls}, using \ToolName with the SEM from
\ifthenelse{\boolean{hideAppendix}}{the extended paper}{\Cref{sec:appendix-JMM}});
the model proposed in~\cite{jam21} (\jamtwentyone, using \herd);
and the multi-execution model implemented in \JavaSim~\cite{ftfjp/MP02}
which aims to realize \jls.
The programs (litmus tests) used during testing come from various sources:
(1) CTC~\cite{causality-test-cases} are the well known ``causality test cases''
designed for evaluation of multi-execution models;
(2) jcstress~\cite{jcstress}
are programs employed by \jcstress
for testing Java implementations' conformance
to the OpenJDK developers understanding or opinion of Java~5 and \VarHandle semantics;
(3) S\&A from~\cite{ecoop/SA08}; and
(4) MT from Manson’s thesis~\cite{manson-thesis}.
Tests corresponding to features outside our scope,
such as final fields
(unsupported in the Java syntax of \herd  and orthogonal to \VarHandle)
and programs with loops
are excluded.
We converted all programs into the \texttt{.litmus} format of \herd 
and extended it with (required and forbidden) behavior assertions.
A behavior is marked required (resp.\ forbidden)
if expert documentation specifies it should be allowed (resp.\ forbidden),
or if the Java language specification~\cite{jmm-jls} applies (i.e., the program is in Java~5) and allows (resp.\ forbids) it.
In addition, we include \numOtherTests further programs with no agreed-upon behavior. 
Interesting results on these programs thus need to be individually looked at,
which we do for two such programs below.

\begin{table}[t]
	\centering
	\caption{Test results of Java memory models (given as pass / fail / unsupported).}
	\label{tab:sc1}
	\smallskip
	\begin{tabular}{l c p{1em} ccccc p{1em} ccccc p{1em} ccccc}
		\toprule
		& \# tests && \multicolumn{5}{c}{\ToolName-\jls} && \multicolumn{5}{c}{\herd-\jamtwentyone} && \multicolumn{5}{c}{\JavaSim} \\ \midrule
		CTC 1--16             &   13 &&  13 &/& 0 &/& 0 &&  8 &/& {\color{red}5} &/& 0 && 11 &/&             0  &/& 2 \\
		CTC 17, 18            &   2  &&   2 &/& 0 &/& 0 &&  2 &/&             0  &/& 0 &&  0 &/& {\color{red}2} &/& 0 \\
		jcstress - Java~5     &   27 &&  27 &/& 0 &/& 0 && 27 &/&             0  &/& 0 && 27 &/&             0  &/& 0 \\
		jcstress - \VarHandle &   28 &&     & &-- & &   && 24 &/& {\color{red}1} &/& 3 &&    & &            --  & &   \\
		S\&A                  &   6  &&   6 &/& 0 &/& 0 &&  4 &/& {\color{red}2} &/& 0 &&  4 &/& 2\tablefootnote{For these incorrect results \JavaSim already gave a warning.} &/& 0 \\
		MT                    &   24 &&  24 &/& 0 &/& 0 && 16 &/& {\color{red}6} &/& 2 && 18 &/& {\color{red}4} &/& 2 \\
		\midrule
		\textbf{total}        &  100 &&  72 &/& 0 &/& 0 && 81 &/& {\color{red}14}&/& 5 && 60 &/& {\color{red}8} &/& 4 \\
		\bottomrule
	\end{tabular}
\end{table}	

\inlineheadingbf{Scenario \circled{1}}
In this scenario, we check whether a proposed Java memory model
allows all the required and forbids all the forbidden behaviors as described within a litmus test.
In this case, we record it as ``pass''; in all other cases (a required behavior is not allowed, or a forbidden behavior is possible)
we record it as ``fail''.
Furthermore, we have outcome ``unsupported'' when tools do not support the input or crash on it.
For example, \herd and \JavaSim do not support the bitwise \texttt{or} operator and
crash on programs containing it.
We do not report timeouts,
since all tests finish within a minute
(most within a second, \ifthenelse{\boolean{hideAppendix}}{details in the extended paper}{see also \Cref{sec:results-full}}).
This performance is comparable to our experience with other weak-memory tools:
fast enough for iterative model and test development. 

\Cref{tab:sc1} shows the summarized results of our experiments concerning scenario~\circled{1}.
The full results are in \ifthenelse{\boolean{hideAppendix}}{the extended paper}{Appendix~\ref{sec:results-full}}.
We see that for two of the three proposed memory models we are able
to find programs showing differences in the expected behavior and that of the memory model. For \jls and \JavaSim, we have elided experiments with
litmus tests using the \VarHandle API, because the language specification was developed prior to this API.
An interesting result is that \JavaSim, designed to implement the Java language specification, actually differs from \jls on 8 litmus tests.
Furthermore, we validated that \ToolName, given the \jls SEM, implements \jls.

\inlineheadingbf{Scenario \circled{2}} In these experiments, we check whether
the behavior of a litmus test compiled to x86 with our non-optimizing compiler
only shows behavior allowed by a memory model. On our test suite of scenario \circled{1},
\ToolName finds no differences. We thus also ran our tool
on the additional programs, revealing unsoundness of \jamtwentyone with respect to compilation.

\begin{figure}[t]
	\begin{subfigure}{\textwidth}
		\centering
		\begin{tabular}[b]{@{}c@{}}
			$\cloc{x}=\cloc{y}=\cloc{z}=0$ \\
			\smallskip
			$\begin{array}{@{}l@{~}||@{~}l@{~}||@{~}l@{}}
				\begin{array}[t]{l}
					\writeInst{\cloc{x}}{\creg{2}}{\lRel}; \\
					\readInst{\creg{a}}{\cloc{x}}{\lAcq}; \itreads{2} \\
					\readInst{\creg{b}}{\cloc{z}}{\lVol};\, \itreads{0} \\
					\readInst{\creg{c}}{\cloc{y}}{\lVol}\;\; \itreads{0} \\
				\end{array}
				&
				\begin{array}[t]{l}
					\writeInst{\cloc{y}}{\creg{2}}{\lVol}; \\
					\writeInst{\cloc{x}}{\creg{1}}{\lVol} \\
				\end{array}
				&
				\begin{array}[t]{l}
					\readInst{\creg{d}}{\cloc{x}}{\lOpq}; \itreads{1} \\
					\readInst{\creg{e}}{\cloc{x}}{\lOpq}\;\, \itreads{2} \\
				\end{array}
			\end{array}$
		\end{tabular}
		\vspace*{-.4cm}
		\caption{}\label{fig:comp-err-source}
		\vspace*{.2cm}
	\end{subfigure}
	\begin{subfigure}{\textwidth}
		\centering
		\begin{tabular}[b]{@{}c@{}}
			$\cloc{x}=\cloc{y}=\cloc{z}=0$ \\
			\smallskip
			$\begin{array}{@{}l@{~}||@{~}l@{~}||@{~}l@{}}
				\begin{array}[t]{l}
					\mathtt{mov}\ [\cloc{x}],\, 2 \\
					\mathtt{mov}\ \creg{EAX},\, [\cloc{x}] \itreads[;]{2} \\
					\mathtt{mov}\ \creg{EBX},\, [\cloc{z}] \itreads[;]{0} \\
					\mathtt{mov}\ \creg{ECX},\, [\cloc{y}] \itreads[;]{0} \\
				\end{array}
				&
				\begin{array}[t]{l}
					\mathtt{mov}\ [\cloc{y}],\, 2 \\
					\mathtt{mfence} \\
					\mathtt{mov}\ [\cloc{x}],\, 1 \\
					\mathtt{mfence} \\
				\end{array}
				&
				\begin{array}[t]{l}
					\mathtt{mov}\ \creg{EAX},\, [\cloc{x}] \itreads[;]{1} \\
					\mathtt{mov}\ \creg{EBX},\, [\cloc{x}] \itreads[;]{2} \\
				\end{array}
			\end{array}$
		\end{tabular}
		\vspace*{-.2cm}
		\caption{}\label{fig:comp-err-target}
		\vspace*{-.2cm}
	\end{subfigure}
	\caption{
		A program (\subref{fig:comp-err-source})
		which when compiled to x86 (\subref{fig:comp-err-target})
		using the compilation scheme of~\cite{jam21}
		has a behavior (values in comments, $\itreads{}$ and $\itreads[;]{}$) that the source program does not allow
		according to their memory model~\jamtwentyone.
	}\label{fig:comp-err}
	\vspace*{-.2cm}
\end{figure}

More specifically, one of the programs in our additional test set is
taken from the \jamtwentyone article (Section~{G.2.3.1} of the appendix of~\cite{jam21}).
For this program, \herd (on the compiled program) calculates a behavior
which \jamtwentyone does not allow. This is slightly surprising since
our non-optimizing compiler uses exactly the compilation scheme given in~\cite{jam21}
for which the authors of~\cite{jam21} have proven correctness.
We simplified the program to the core necessary to see the unsound compilation.
The simplified program is given in Figure~\ref{fig:comp-err}.

Obtaining such a result for a Java memory model might
not necessarily mean that the memory model is broken,
it could also be an unsound compilation scheme.
Nevertheless, during the development of memory models such results
are an indispensable aid for design which is what \ToolName is meant for.

\inlineheadingbf{Scenario \circled{3}} For this scenario,
we use \jcstress and a standard Java compiler.
Our experiments basically give the same results as for scenario \circled{2}:
on our base test suite no differences are observable; on the extended set we find a number of programs
with differing behavior.

One program with differences is again the one in \Cref{fig:comp-err}.
Using our conversion to \jcstress,
we were able to show that the result is also observed
on x86 with OpenJDK 24 in about one in 18 thousand executions.
Another interesting example where \jamtwentyone prevents a behavior
that \jcstress can observe
is the {SB+rfis} example from~\cite[Section 4, Remark 4]{pldi/LVKHD17}:

\noindent\hspace*{\fill}
\begin{minipage}{.51\textwidth}
	\centering
	\begin{tabular}[b]{@{}c@{}}
		$\cloc{x}=\cloc{y}=0$ \\
		\smallskip
		$\begin{array}{@{}l@{~}||@{~}l@{}}
			\begin{array}[t]{l}
				\writeInst{\cloc{x}}{\creg{1}}{\lPln}; \\
				\readInst{\creg{a}}{\cloc{x}}{\lVol}; \itreads{1} \\
				\readInst{\creg{b}}{\cloc{y}}{\lVol}\;\, \itreads{0} \\
			\end{array}
			&
			\begin{array}[t]{l}
				\writeInst{\cloc{y}}{\creg{1}}{\lPln}; \\
				\readInst{\creg{c}}{\cloc{y}}{\lVol}; \itreads{1} \\
				\readInst{\creg{d}}{\cloc{x}}{\lVol}\;\, \itreads{0} \\
			\end{array}
		\end{array}$
	\end{tabular}
\end{minipage}
\hspace*{\fill}
\begin{minipage}[b]{0.27\textwidth}
	\centering
	\begin{tabular}[c]{c@{\hspace{2pt}}c@{\hspace{2pt}}c@{\hspace{2pt}}c@{\hspace{8pt}} r@{\hspace{8pt}}c}
		$\creg{a}$ & $\creg{b}$ & $\creg{c}$ & $\creg{d}$ & samples & $\jamtwentyone$ \\
		\hline
		$1$ & $0$ & $1$ & $0$ & $  324 \, \mathrm{M}$ & \forbidden \\
		$1$ & $0$ & $1$ & $1$ & $1,240 \, \mathrm{M}$ &   \allowed \\
		$1$ & $1$ & $1$ & $0$ & $1,135 \, \mathrm{M}$ &   \allowed \\
		$1$ & $1$ & $1$ & $1$ & $   37 \, \mathrm{M}$ &   \allowed \\
	\end{tabular}
\end{minipage}
\hspace*{\fill}

\noindent 
On the left is the program;
on the right the register values $\creg{a}$--$\creg{d}$ observed by \jcstress.
The result forbidden by \jamtwentyone occurs in 324 million samples.
This is particularly interesting
as it appears to be a counterexample
to Lea’s explanation~\cite{j9mm} of volatile semantics:
“Volatile mode accesses are totally ordered.”

We interpret this statement as saying that
the total order on volatile mode accesses is included in the antecedence relation
defined in\cite{j9mm}.
Otherwise, this total order would remain unnamed and play no role in the semantics.
Under this interpretation, the forbidden outcome necessarily induces a coherence violation.
Without loss of generality (the two threads are symmetric),
assume that the volatile read
$\lR^{\lVol}_1(\cloc{x},\creg{a})$
precedes the volatile read
$\lR^{\lVol}_2(\cloc{y},\creg{c})$
in the total order on volatile accesses. Then we obtain the following cycle in the antecedence relation:
$\lW^{\lPln}_1(\cloc{x},\creg{1})
\xrightarrow{\textsf{antec.}} \lR^{\lVol}_1(\cloc{x},\creg{a})
\xrightarrow{\textsf{antec.}} \lR^{\lVol}_2(\cloc{y},\creg{c})
\xrightarrow{\textsf{antec.}} \lR^{\lVol}_2(\cloc{x},\creg{d})
\xrightarrow{\textsf{antidep.}} \lW^{\lPln}_1(\cloc{x},\creg{1})$.
This cycle contradicts the acyclicity of antecedence, and therefore the outcome 
$(\creg{a}, \creg{b}, \creg{c}, \creg{d})=(1,0,1,0)$ is forbidden under this reading of~\cite{j9mm}.

\section{Conclusion}

In this paper, we have introduced \ToolName,
a tool for testing Java memory models
against expected semantics
and their compatibility with compilation schemes and compiler implementations.
For its multi-execution model support,
\ToolName builds on the concept of {\em symbolic execution graphs}
which allows for an efficient representation
of infinitely many potential candidates of execution graphs.
The experimental evaluation shows that \ToolName can be a valuable aid
in evaluating how JMMs match the mentioned desiderata.
In the future, we intend to extend \ToolName
with support for additional features,
e.g., final fields
and compilation to {RISC-V}.
Additionally
-- as an alternative to the sampling approach of \jcstress --
we could also statically analyze the machine code generated by the JIT compiler.
Finally, we believe that \ToolName could play an important role
in supporting the development of an updated JMM
compatible with \jls and the full \VarHandle API.

\paragraph{Data availability statement.}
The models, tools, and scripts
to reproduce our experimental evaluation
are archived and available at~\cite{artefact}.

\bibliographystyle{splncs04}
\bibliography{references,main}
\ifthenelse{\boolean{hideAppendix}}{}{
	\newpage
	\appendix
	\section{The Java language specification's SEM} \label{sec:appendix-JMM}

\begin{lstlisting}[language=cat,label=lst:jls-cat,captionpos=b,caption=\jls SEM \cat model.]
"Java Language Specification"

let volatile = V
let write = W
let read = R

let sync-act = (volatile & write) | (volatile & read)
with so from linearisations(sync-act, po)
let so-loc = so & loc

let first-action = ~range(po) \ IW

let sw =
    [volatile & write] ; (so-loc) ; [volatile & read]
  | IW * first-action

let hb = ( po | sw )*
let hb-loc = hb & loc

irreflexive rf;hb
irreflexive [W];hb-loc;rf^-1;(hb-loc\id)
irreflexive rf;so
irreflexive [W];so-loc;rf^-1;(so-loc\id)

show po, rf, so, sw, hb
\end{lstlisting}

In Listing~\ref{lst:jls-cat}, we give a \cat model of the happens-before memory model
of the Java Language specification~\cite{jmm-jls}.
The happens-before model is the single-execution part of the language specification.
The given listing is a straightforward copy from the specification,
except for one minor difference:
The specification states that
``The happens-before order [...] must be a valid partial order:
\emph{reflexive}, transitive and antisymmetric''
and that
``A set of actions A\footnote{Here, we call these $\Events$.} is happens-before consistent
if for all reads $r$ in $A$, where $W(r)$ is the write action seen by $r$,
it is not the case that [...] there exists a write $w$ in $A$
such that $w.v = r.v$ and $hb(W(r), w)$ and $hb(w, r)$.''
If both hold at the same time then we get trivial violations of the consistency predicate for
single-threaded programs that must be acceptable.
Take for example the single threaded program
that writes $1$ to $\cloc{x}$ (call the write~$w'$)
and then reads from $\cloc{x}$ (call the read~$r$).
Obviously this program must read $1$ (the only legal reads-from relation has $W(r) = w'$
and thus $w'.v = r.v = 1$).
But now there exists a write $w = w'$
with $w.v = r.v = 1$ and $hb(W(r), w)$ (by $W(r) = w' = w$ and reflexivity of $hb$) and $hb(w, r)$.
Thus, either $hb$ must be irreflexive, or in the property $w$ and $w'$ must be different.
We decided to go with the second option, because $w$ is called an ``interleaving'' write,
and it does not make sense to us for a write to interleave with itself.

	\section{Full \VarHandle to x86 Compilation Scheme}\label{sec:x86-compilation-scheme}

\begin{table}[t]
	\centering
	\caption{
		A proposed Java \VarHandle to x86 compilation scheme.
	}\label{tbl:x86-compilation-scheme}
	 \smallskip
	\begin{tabular}{lcl}
		Function & & Implementaion \\

		$r_\mathit{dst} = X.\mathtt{get}()$         & $\leadsto$ & $\mathtt{mov}\ r_\mathit{dst},\, [x]$ \\
		$r_\mathit{dst} = X.\mathtt{getOpaque}()$   & $\leadsto$ & $\mathtt{mov}\ r_\mathit{dst},\, [x]$ \\
		$r_\mathit{dst} = X.\mathtt{getAcquire}()$  & $\leadsto$ & $\mathtt{mov}\ r_\mathit{dst},\, [x]$ \\
		$r_\mathit{dst} = X.\mathtt{getVolatile}()$ & $\leadsto$ & $\mathtt{mov}\ r_\mathit{dst},\, [x]$ \\

		$X.\mathtt{set}(r_\mathit{arg})$         & $\leadsto$ & $\mathtt{mov}\ [x],\, r_\mathit{arg}$ \\
		$X.\mathtt{setOpaque}(r_\mathit{arg})$   & $\leadsto$ & $\mathtt{mov}\ [x],\, r_\mathit{arg}$ \\
		$X.\mathtt{setRelease}(r_\mathit{arg})$  & $\leadsto$ & $\mathtt{mov}\ [x],\, r_\mathit{arg}$ \\
		$X.\mathtt{setVolatile}(r_\mathit{arg})$ & $\leadsto$ & $\mathtt{mov}\ [x],\, r_\mathit{arg} ;\; \mathtt{mfence}$ \\

		$\mathtt{fullFence}()$       & $\leadsto$ & $\mathtt{mfence}$ \\
		$\mathtt{acquireFence}()$    & $\leadsto$ & $\mathtt{nop}$ \\
		$\mathtt{releaseFence}()$    & $\leadsto$ & $\mathtt{nop}$ \\
		$\mathtt{loadLoadFence}()$   & $\leadsto$ & $\mathtt{nop}$ \\
		$\mathtt{storeStoreFence}()$ & $\leadsto$ & $\mathtt{nop}$ \\

		\rule{0pt}{2.5ex}$r_\mathit{dst} = X.\mathtt{getAndSet}(r_\mathit{arg})$ & $\leadsto$ &
			\makecell[tl]{
				$\mathtt{mov}\ r_\mathit{tmp},\, r_\mathit{arg}$ \\
				$\mathtt{xchg}\ [x],\, r_\mathit{tmp}$ \\
				$\mathtt{mov}\ r_\mathit{dst},\, r_\mathit{tmp}$ \\
			} \\
		\rule{0pt}{2.5ex}$r_\mathit{dst} = X.\mathtt{getAndAdd}(r_\mathit{arg})$ & $\leadsto$ &
			\makecell[tl]{
				$\mathtt{mov}\ r_\mathit{tmp},\, r_\mathit{arg}$ \\
				$\mathtt{lock}\ \mathtt{xadd}\ [x],\, r_\mathit{tmp}$ \\
				$\mathtt{mov}\ r_\mathit{dst},\, r_\mathit{tmp}$ \\
			} \\
		\rule{0pt}{2.5ex}$r_\mathit{dst} = X.\mathtt{compareAndSet}(r_\mathit{exp},\, r_\mathit{new})$ & $\leadsto$ &
			\makecell[tl]{
				$\mathtt{mov}\ \mathtt{EAX},\, r_\mathit{exp}$ \\
				$\mathtt{mov}\ r_\mathit{tmp},\, r_\mathit{new}$ \\
				$\mathtt{lock}\ \mathtt{cmpxchg}\ [x],\, r_\mathit{tmp}$ \\
				$\mathtt{sete}\ r_\mathit{dst}$ \\
			} \\
		\rule{0pt}{2.5ex}$r_\mathit{dst} = X.\mathtt{compareAndExchange}(r_\mathit{exp},\, r_\mathit{new})$ & $\leadsto$ &
			\makecell[tl] {
				$\mathtt{mov}\ \mathtt{EAX},\, r_\mathit{exp}$ \\
				$\mathtt{mov}\ r_\mathit{tmp},\, r_\mathit{new}$ \\
				$\mathtt{lock}\ \mathtt{cmpxchg}\ [x],\, r_\mathit{tmp}$ \\
				$\mathtt{mov}\ r_\mathit{dst},\, \mathtt{EAX}$ \\
			} \\
	\end{tabular}
	
\end{table}

Table~\ref{tbl:x86-compilation-scheme}
shows a possible compilation scheme.
The compilation scheme of~\cite[Appendix F.0.2]{jam21} is the same as ours on the instructions common to both schemes.
For simplicity this table assumes that all values are in registers that must not be overwritten.
If a value is no longer needed after a function call,
or if it is a constant,
some move instructions could be omitted.
In the table we have local variables $r$ on the function side
that correspond to registers on the implementation side.
We give these the same name in the table.
In a later register allocation phase we will assign these names to registers of the target platform.
For a reading function,  $r_\mathit{dst}$
is the local variable where a read places the value it got from memory.
For a writing function,  $r_\mathit{arg}$
is the local variable containing the value written to memory.
For a compare-and-update function, $r_\mathit{exp}$
is the expected value
and $r_\mathit{new}$
is the value written if the compare succeeds.

The \VarHandle functions not defined in the table are implemented in the following way:
Other volatile get-and-update functions
(\texttt{get\-And\-Bitwise\-And},
\texttt{get\-And\-Bitwise\-Or},
\texttt{get\-And\-Bitwise\-Xor})
are implemented as compare-and-exchange loops.
The release and acquire variants of
the get-and-update functions
(\texttt{get\-And\-Set\-Ac\-quire},
\texttt{get\-And\-Set\-Re\-lease},
\texttt{get\-And\-Add\-Ac\-quire},
\texttt{get\-And\-Add\-Re\-lease},
\texttt{get\-And\-Bit\-wise\-And\-Ac\-quire},
\texttt{get\-And\-Bit\-wise\-And\-Re\-lease},
\texttt{get\-And\-Bit\-wise\-Or\-Ac\-quire},
\texttt{get\-And\-Bit\-wise\-Or\-Re\-lease},
\texttt{get\-And\-Bit\-wise\-Xor\-Ac\-quire},
\texttt{get\-And\-Bit\-wise\-Xor\-Re\-lease}),
are implemented as their volatile (no suffix) counterpart.
The weak compare-and-set functions
(\texttt{weak\-Compare\-And\-Set\-Plain},
\texttt{weak\-Compare\-And\-Set\-Ac\-quire},
\texttt{weak\-Compare\-And\-Set\-Re\-lease},
\texttt{weak\-Compare\-And\-Set})
are implemented as \texttt{compare\-And\-Set}.

	\section{Results}\label{sec:results-full}

Tables
\ref{tbl:sc1-ctc},
\ref{tbl:sc1-mt},
\ref{tbl:sc1-sa},
\ref{tbl:sc1-jcstress-coherence},
\ref{tbl:sc1-jcstress-causality},
\ref{tbl:sc1-jcstress-consensus} and
\ref{tbl:sc1-jcstress-other}
show the individual test results that are summarized in Table~\ref{tab:sc1}.
For each test we list the expected result.
We describe in Section~\ref{sec:evaluation} how we decide on the expected result.
Here, we write \Sometimes for required and \Never for forbidden.
We list for each test the verdict of the tool.
We write \Sometimes when the model and tool combination allows the result,
\Never if it is not allowed,
and \Crash if the tool does not support the input or crashed,
We highlight a mismatch in {\color{red!90!black}red},
unless the tool also produced a warning,
then we highlight the mismatch in {\color{orange!90!black}orange}.
We also list the wall-clock time spent on the calculation.
For \JavaSim we include the time that \ToolName takes
to convert the litmus test into \JavaSim's format
and to parse and interpret \JavaSim's output.
We observe that in almost all cases \ToolName with \jls
is slightly slower that \JavaSim,
but within the same order of magnitude.
The tests were executed on a 22 Core Intel(R) Core(TM) Ultra 7 165H Laptop CPU with 32 GB RAM.

\begin{table}
	\centering
	\caption{Results for Causality Test Cases~\cite{causality-test-cases}.}\label{tbl:sc1-ctc}
	\smallskip
	\begin{tabular}{l p{1em} c p{1em} cc p{1em} cc p{1em} cc}
		\toprule
		Test && \multicolumn{1}{c}{Exp.} && \multicolumn{2}{c}{\ToolName-\jls} && \multicolumn{2}{c}{\herd-\jamtwentyone} && \multicolumn{2}{c}{\JavaSim} \\
		\midrule
		Test 1  && \Sometimes && \Sometimes &  0.62s &&             \Sometimes  & 0.01s &&             \Sometimes  & 0.51s \\
		Test 2  && \Sometimes && \Sometimes &  0.82s &&             \Sometimes  & 0.01s &&             \Sometimes  & 0.49 s \\
		Test 3  && \Sometimes && \Sometimes &  0.98s &&             \Sometimes  & 0.01s &&             \Sometimes  & 2.53 s \\
		Test 4  &&     \Never &&     \Never &  0.64s &&                 \Never  & 0.01s &&             \Sometimes  & 0.38 s \\
		Test 5  &&     \Never &&     \Never & 23.78s &&                 \Never  & 0.01s &&             \Sometimes  & 0.68 s \\
		Test 6  && \Sometimes && \Sometimes &  0.70s &&             \Sometimes  & 0.01s &&             \Sometimes  & 0.40 s \\
		Test 7  && \Sometimes && \Sometimes &  0.94s &&             \Sometimes  & 0.01s &&             \Sometimes  & 0.47 s \\
		Test 8  && \Sometimes && \Sometimes &  0.65s && {\color{red}    \Never} & 0.01s &&                 \Crash  & 0.38 s \\
		Test 9  && \Sometimes && \Sometimes &  0.77s && {\color{red}    \Never} & 0.01s &&                 \Crash  & 0.46 s \\
		Test 10 &&     \Never &&     \Never &  0.67s && {\color{red}\Sometimes} & 0.01s &&             \Sometimes  & 0.68 s \\
		Test 11 && \Sometimes && \Sometimes &  0.74s &&             \Sometimes  & 0.02s &&             \Sometimes  & 0.54 s \\
		Test 13 &&     \Never &&     \Never &  0.39s && {\color{red}\Sometimes} & 0.01s &&             \Sometimes  & 0.37 s \\
		Test 16 && \Sometimes && \Sometimes &  0.61s && {\color{red}    \Never} & 0.01s &&             \Sometimes  & 0.48 s \\
		Test 17 &&     \Never &&     \Never &  0.51s &&                 \Never  & 0.01s && {\color{red}\Sometimes} & 0.50 s \\
		Test 18 &&     \Never &&     \Never &  0.61s &&                 \Never  & 0.01s && {\color{red}\Sometimes} & 0.49 s \\
		\bottomrule
	\end{tabular}
\end{table}

\begin{table}
	\centering
	\caption{Results for tests from Manson's thesis~\cite{manson-thesis}.}\label{tbl:sc1-mt}
	\smallskip
	\begin{tabular}{l p{1em} c p{1em} cc p{1em} cc p{1em} cc}
		\toprule
		Test && \multicolumn{1}{c}{Exp.} && \multicolumn{2}{c}{\ToolName-\jls} && \multicolumn{2}{c}{\herd-\jamtwentyone} && \multicolumn{2}{c}{\JavaSim} \\
		\midrule
		Fig.~1.2  &&     \Never &&     \Never &  0.65s &&                 \Never  & 0.01s &&                 \Never  & 0.44s \\
		Fig.~1.3  && \Sometimes && \Sometimes &  0.59s &&                 \Crash  & 0.01s &&                 \Crash  & 0.55s \\
		Fig.~2.1  && \Sometimes && \Sometimes &  0.58s &&             \Sometimes  & 0.01s &&             \Sometimes  & 0.46s \\
		Fig.~2.2  && \Sometimes && \Sometimes &  0.72s &&             \Sometimes  & 0.01s &&             \Sometimes  & 0.47s \\
		Fig.~2.5  && \Sometimes && \Sometimes &  0.72s &&             \Sometimes  & 0.01s &&             \Sometimes  & 0.58s \\
		Fig.~3.1  &&     \Never &&     \Never &  0.55s && {\color{red}\Sometimes} & 0.01s &&                 \Never  & 0.50s \\
		Fig.~3.3a && \Sometimes && \Sometimes &  0.51s &&             \Sometimes  & 0.01s &&             \Sometimes  & 0.40s \\
		Fig.~3.3b && \Sometimes && \Sometimes &  0.59s &&                 \Crash  & 0.01s &&                 \Crash  & 0.35s \\
		Fig.~3.5  && \Sometimes && \Sometimes &  0.59s &&             \Sometimes  & 0.01s &&             \Sometimes  & 0.45s \\
		Fig.~3.6  && \Sometimes && \Sometimes &  0.68s && {\color{red}    \Never} & 0.01s &&             \Sometimes  & 0.34s \\
		Fig.~3.9  &&     \Never &&     \Never &  1.35s &&                 \Never  & 0.01s &&                 \Never  & 0.52s \\
		Fig.~3.10 &&     \Never &&     \Never &  0.65s &&                 \Never  & 0.02s &&                 \Never  & 0.50s \\
		Fig.~4.5  &&     \Never &&     \Never & 24.84s &&                 \Never  & 0.01s &&                 \Never  & 0.65s \\
		Fig.~4.6  &&     \Never &&     \Never &  0.72s &&                 \Never  & 0.01s &&                 \Never  & 0.48s \\
		Fig.~4.7  &&     \Never &&     \Never &  0.77s &&                 \Never  & 0.01s && {\color{red}\Sometimes} & 0.49s \\
		Fig.~4.8  &&     \Never &&     \Never &  0.65s && {\color{red}\Sometimes} & 0.01s &&                 \Never  & 0.48s \\
		Fig.~4.9  &&     \Never &&     \Never &  0.76s && {\color{red}\Sometimes} & 0.01s &&                 \Never  & 3.57s \\
		Fig.~4.10 &&     \Never &&     \Never &  0.74s && {\color{red}\Sometimes} & 0.01s && {\color{red}\Sometimes} & 0.53s \\
		Fig.~4.11 &&     \Never &&     \Never &  0.87s &&                 \Never  & 0.01s &&                 \Never  & 0.53s \\
		Fig.~4.12 &&     \Never &&     \Never &  0.69s &&                 \Never  & 0.01s && {\color{red}\Sometimes} & 0.45s \\
		Fig.~5.1  && \Sometimes && \Sometimes &  0.58s &&             \Sometimes  & 0.01s &&             \Sometimes  & 0.52s \\
		Fig.~8.1  && \Sometimes && \Sometimes &  0.76s &&             \Sometimes  & 0.01s && {\color{red}    \Never} & 0.42s \\
		Fig.~8.2  && \Sometimes && \Sometimes &  0.49s && {\color{red}    \Never} & 0.01s &&             \Sometimes  & 0.52s \\
		Fig.~8.3  && \Sometimes && \Sometimes &  0.63s &&             \Sometimes  & 0.01s &&             \Sometimes  & 0.63s \\
	\end{tabular}
\end{table}

\begin{table}
	\centering
	\caption{Results for tests by {\v{S}}ev{\v{c}}{\'{i}}k and Aspinall~\cite{ecoop/SA08}.}\label{tbl:sc1-sa}
	\smallskip
	\begin{tabular}{l p{1em} c p{1em} cc p{1em} cc p{1em} cc}
		\toprule
		Test && \multicolumn{1}{c}{Exp.} && \multicolumn{2}{c}{\ToolName-\jls} && \multicolumn{2}{c}{\herd-\jamtwentyone} && \multicolumn{2}{c}{\JavaSim} \\
		\midrule
		Read intro. source &&     \Never &&      \Never & 0.65s &&                 \Never  & 0.01s &&                 \Never  & 0.43s \\
		Read intro. target && \Sometimes &&  \Sometimes & 0.86s && {\color{red}    \Never} & 0.01s && {\color{orange} \Never} & 0.59s \\
		Read elim. source  &&     \Never &&      \Never & 0.61s &&                 \Never  & 0.01s &&                 \Never  & 0.36s \\
		Read elim. target  && \Sometimes &&  \Sometimes & 0.91s && {\color{red}    \Never} & 0.01s && {\color{orange} \Never} & 0.43s \\
		Fig.~5 source      &&     \Never &&      \Never & 0.65s &&                 \Never  & 0.01s &&                 \Never  & 0.60s \\
		Fig.~5 target      && \Sometimes &&  \Sometimes & 0.61s &&             \Sometimes  & 0.01s &&             \Sometimes  & 0.46s \\
		\bottomrule
	\end{tabular}
\end{table}

\begin{table}
	\centering
	\caption{Results for coherence tests from \jcstress~\cite{jcstress}.}\label{tbl:sc1-jcstress-coherence}
	\smallskip
	\begin{tabular}{l p{1em} c p{1em} cc p{1em} cc p{1em} cc}
		\toprule
		Test && \multicolumn{1}{c}{Exp.} && \multicolumn{2}{c}{\ToolName-\jls} && \multicolumn{2}{c}{\herd-\jamtwentyone} && \multicolumn{2}{c}{\JavaSim} \\
		\midrule
		Plain    0 0 && \Sometimes && \Sometimes & 0.55s &&             \Sometimes  & 0.01s && \Sometimes & 0.44s \\
		Plain    0 1 && \Sometimes && \Sometimes & 0.64s &&             \Sometimes  & 0.01s && \Sometimes & 0.46s \\
		Plain    1 0 && \Sometimes && \Sometimes & 0.53s &&             \Sometimes  & 0.01s && \Sometimes & 0.43s \\
		Plain    1 1 && \Sometimes && \Sometimes & 0.58s &&             \Sometimes  & 0.01s && \Sometimes & 0.42s \\
		Opaque   0 0 && \Sometimes &&         -- &       &&             \Sometimes  & 0.01s &&         -- &       \\
		Opaque   0 1 && \Sometimes &&         -- &       &&             \Sometimes  & 0.01s &&         -- &       \\
		Opaque   1 0 &&     \Never &&         -- &       && {\color{red}\Sometimes} & 0.01s &&         -- &       \\
		Opaque   1 1 && \Sometimes &&         -- &       &&             \Sometimes  & 0.01s &&         -- &       \\
		Volatile 0 0 && \Sometimes && \Sometimes & 0.52s &&             \Sometimes  & 0.01s && \Sometimes & 0.44s \\
		Volatile 0 1 && \Sometimes && \Sometimes & 0.58s &&             \Sometimes  & 0.01s && \Sometimes & 0.43s \\
		Volatile 1 0 &&     \Never &&     \Never & 0.52s &&                 \Never  & 0.01s &&     \Never & 0.42s \\
		Volatile 1 1 && \Sometimes && \Sometimes & 0.46s &&             \Sometimes  & 0.01s && \Sometimes & 0.33s \\
		\bottomrule
	\end{tabular}
\end{table}

\begin{table}
	\centering
	\caption{Results for causality (message passing) tests from \jcstress~\cite{jcstress}.}\label{tbl:sc1-jcstress-causality}
	\smallskip
	\begin{tabular}{l p{1em} c p{1em} cc p{1em} cc p{1em} cc}
		\toprule
		Test && \multicolumn{1}{c}{Exp.} && \multicolumn{2}{c}{\ToolName-\jls} && \multicolumn{2}{c}{\herd-\jamtwentyone} && \multicolumn{2}{c}{\JavaSim} \\
		\midrule
		Plain      0 0 && \Sometimes && \Sometimes & 0.50s &&             \Sometimes  & 0.01s && \Sometimes & 0.46s \\
		Plain      0 1 && \Sometimes && \Sometimes & 0.49s &&             \Sometimes  & 0.01s && \Sometimes & 0.55s \\
		Plain      1 0 && \Sometimes && \Sometimes & 0.61s &&             \Sometimes  & 0.01s && \Sometimes & 0.43s \\
		Plain      1 1 && \Sometimes && \Sometimes & 0.48s &&             \Sometimes  & 0.01s && \Sometimes & 0.43s \\
		Opaque     0 0 && \Sometimes &&         -- &       &&             \Sometimes  & 0.01s &&         -- &       \\
		Opaque     0 1 && \Sometimes &&         -- &       &&             \Sometimes  & 0.01s &&         -- &       \\
		Opaque     1 0 && \Sometimes &&         -- &       &&             \Sometimes  & 0.01s &&         -- &       \\
		Opaque     1 1 && \Sometimes &&         -- &       &&             \Sometimes  & 0.01s &&         -- &       \\
		Rel./Acq.\ 0 0 && \Sometimes &&         -- &       &&             \Sometimes  & 0.01s &&         -- &       \\
		Rel./Acq.\ 0 1 && \Sometimes &&         -- &       &&             \Sometimes  & 0.01s &&         -- &       \\
		Rel./Acq.\ 1 0 &&     \Never &&         -- &       &&                 \Never  & 0.01s &&         -- &       \\
		Rel./Acq.\ 1 1 && \Sometimes &&         -- &       &&             \Sometimes  & 0.01s &&         -- &       \\
		Volatile   0 0 && \Sometimes && \Sometimes & 0.72s &&             \Sometimes  & 0.01s && \Sometimes & 0.40s \\
		Volatile   0 1 && \Sometimes && \Sometimes & 0.62s &&             \Sometimes  & 0.01s && \Sometimes & 0.44s \\
		Volatile   1 0 &&     \Never &&     \Never & 0.46s &&                 \Never  & 0.01s &&     \Never & 0.43s \\
		Volatile   1 1 && \Sometimes && \Sometimes & 0.68s &&             \Sometimes  & 0.01s && \Sometimes & 0.42s \\
		\bottomrule
	\end{tabular}
\end{table}

\begin{table}
	\centering
	\caption{Results for consensus (store buffering) tests from \jcstress~\cite{jcstress}.}\label{tbl:sc1-jcstress-consensus}
	\smallskip
	\begin{tabular}{l p{1em} c p{1em} cc p{1em} cc p{1em} cc}
		\toprule
		Test && \multicolumn{1}{c}{Exp.} && \multicolumn{2}{c}{\ToolName-\jls} && \multicolumn{2}{c}{\herd-\jamtwentyone} && \multicolumn{2}{c}{\JavaSim} \\
		\midrule
		Plain      0 0 && \Sometimes && \Sometimes & 0.52s &&             \Sometimes  & 0.01s && \Sometimes & 0.33s \\
		Plain      0 1 && \Sometimes && \Sometimes & 0.65s &&             \Sometimes  & 0.01s && \Sometimes & 0.44s \\
		Plain      1 0 && \Sometimes && \Sometimes & 0.55s &&             \Sometimes  & 0.01s && \Sometimes & 0.45s \\
		Plain      1 1 && \Sometimes && \Sometimes & 0.49s &&             \Sometimes  & 0.01s && \Sometimes & 0.51s \\
		Opaque     0 0 && \Sometimes &&         -- &       &&             \Sometimes  & 0.01s &&         -- &       \\
		Opaque     0 1 && \Sometimes &&         -- &       &&             \Sometimes  & 0.01s &&         -- &       \\
		Opaque     1 0 && \Sometimes &&         -- &       &&             \Sometimes  & 0.01s &&         -- &       \\
		Opaque     1 1 && \Sometimes &&         -- &       &&             \Sometimes  & 0.01s &&         -- &       \\
		Rel./Acq.\ 0 0 && \Sometimes &&         -- &       &&             \Sometimes  & 0.01s &&         -- &       \\
		Rel./Acq.\ 0 1 && \Sometimes &&         -- &       &&             \Sometimes  & 0.01s &&         -- &       \\
		Rel./Acq.\ 1 0 && \Sometimes &&         -- &       &&             \Sometimes  & 0.01s &&         -- &       \\
		Rel./Acq.\ 1 1 && \Sometimes &&         -- &       &&             \Sometimes  & 0.01s &&         -- &       \\
		Volatile   0 0 &&     \Never &&     \Never & 0.58s &&                 \Never  & 0.01s &&     \Never & 0.43s \\
		Volatile   0 1 && \Sometimes && \Sometimes & 0.64s &&             \Sometimes  & 0.01s && \Sometimes & 0.45s \\
		Volatile   1 0 && \Sometimes && \Sometimes & 0.61s &&             \Sometimes  & 0.01s && \Sometimes & 0.44s \\
		Volatile   1 1 && \Sometimes && \Sometimes & 0.67s &&             \Sometimes  & 0.01s && \Sometimes & 0.43s \\
		\bottomrule
	\end{tabular}
\end{table}

\begin{table}
	\centering
	\caption{Results for other tests from \jcstress~\cite{jcstress}.}\label{tbl:sc1-jcstress-other}
	\smallskip
	\begin{tabular}{l p{1em} c p{1em} cc p{1em} cc p{1em} cc}
		\toprule
		Test && \multicolumn{1}{c}{Exp.} && \multicolumn{2}{c}{\ToolName-\jls} && \multicolumn{2}{c}{\herd-\jamtwentyone} && \multicolumn{2}{c}{\JavaSim} \\
		\midrule
		\texttt{UnobsVolBarrier}              && \Sometimes && \Sometimes & 1.05s &&             \Sometimes  & 0.01s && \Sometimes & 0.49s \\
		\texttt{vol-not-fence}                && \Sometimes && \Sometimes & 0.95s &&             \Sometimes  & 0.01s && \Sometimes & 0.51s \\
		\texttt{mca-fenced}                   && \Sometimes &&         -- &       &&             \Sometimes  & 0.01s &&         -- &       \\
		\texttt{mca-fully-fenced}             &&     \Never &&         -- &       &&                 \Never  & 0.02s &&         -- &       \\
		\texttt{mca-opaque}                   && \Sometimes &&         -- &       &&             \Sometimes  & 0.01s &&         -- &       \\
		\texttt{non-mca-coherence-1221}       &&     \Never &&         -- &       &&                 \Never  & 0.03s &&         -- &       \\
		\texttt{non-mca-coherence-2112}       &&     \Never &&         -- &       &&                 \Never  & 0.03s &&         -- &       \\
		\texttt{rmw-09-gas-effects-2-cts-cts} && \Sometimes && \Sometimes & 0.96s &&             \Sometimes  & 0.01s && \Sometimes & 0.52s \\
		\texttt{rmw-09-gas-effects-2-cas-cas} && \Sometimes &&         -- &       &&                 \Crash  & 0.01s &&         -- &       \\
		\texttt{rmw-09-gas-effects-3-gts-cas} && \Sometimes &&         -- &       &&                 \Crash  & 0.01s &&         -- &       \\
		\texttt{rmw-09-gas-effects-4-gas-cas} &&     \Never &&         -- &       &&                 \Crash  & 0.01s &&         -- &       \\
		\bottomrule
	\end{tabular}
\end{table}

}

\end{document}